\documentclass{aa}  

\usepackage{graphicx}
\usepackage{txfonts}
\usepackage{placeins}
\usepackage[]{hyperref}
\hypersetup{colorlinks=true,citecolor=blue,linkcolor=black, urlcolor=black, linktoc=all}
\begin{document}

   \title{Deuteration of c-C$_3$H$_2$ towards the pre-stellar core L1544}

   \author{K. Giers
          \inst{1},
          S. Spezzano\inst{1}, 
          F. Alves\inst{1},
          P. Caselli\inst{1},
          E. Redaelli\inst{1},
          O. Sipil\"a\inst{1},
          M. Ben Khalifa\inst{2},
          L. Wiesenfeld\inst{3},
          S. Br\"unken\inst{4},
          L. Bizzocchi\inst{5,6}
          }

   \institute{Max Planck Institute for Extraterrestrial Physics, Giessenbachstrasse 1, D-85748 Garching, Germany\\
              \email{kgiers@mpe.mpg.de}
         \and
             KU Leuven, Department of Chemistry, Celestijnenlaan 200F, B-3001 Leuven, Belgium
         \and 
         Laboratoire Aim\'e-Cotton, B\^atiment 505, CNRS, Universit\'e Paris-Saclay, F-91405 Orsay, France
         \and 
         Radboud University, Institute for Molecules and Materials, FELIX Laboratory, Toernooiveld 7, 6525ED Nijmegen, the Netherlands
         \and 
         Scuola Normale Superiore, Piazza dei Cavalieri 7, 56126 Pisa, Italy
         \and 
         Dipartimento di Chimica “Giacomo Ciamician”, Universit\`a di Bologna, Via Selmi 2, 40126 Bologna, Italy
             }


 
  \abstract
   { In the centre of pre-stellar cores, the deuterium fractionation is enhanced due to the cold temperatures and high densities. Therefore, the chemistry of deuterated molecules can be used to probe the evolution and the kinematics in the earliest stages of star formation.}
   {We analyse emission maps of cyclopropenylidene, c-C$_3$H$_2$, to study the distribution of the deuteration throughout the prototypical pre-stellar core L1544.}
   {We use single-dish observations of c-C$_3$H$_2$, c-H$^{13}$CC$_2$H, c-C$_3$HD, and c-C$_3$D$_2$ towards the pre-stellar core L1544, performed at the IRAM 30\,m telescope. We derive the column density and deuterium fraction maps, and compare these observations with non-LTE radiative transfer simulations. }
   {The highest deuterium fractions are found close to the dust peak at the centre of L1544, where the increased abundance of H$_2$D$^+$ ions drives the deuteration process. The peak values are N(c-C$_3$HD)/N(c-C$_3$H$_2)=0.17\pm0.01$, N(c-C$_3$D$_2$)/N(c-C$_3$H$_2)=0.025\pm0.003$ and N(c-C$_3$D$_2$)/N(c-C$_3$HD$)=0.16\pm0.03$, which is consistent with previous single point observations. The distributions of c-C$_3$HD and c-C$_3$D$_2$ indicate that the deuterated forms of c-C$_3$H$_2$ in fact trace the dust peak and not the c-C$_3$H$_2$ peak. 
   }
   {The N(c-C$_3$D$_2$)/N(c-C$_3$HD) map confirms that the process of deuteration is more efficient towards the centre of the core and demonstrates that carbon-chain molecules are still present at high densities. This is likely caused by an increased abundance of He$^+$ ions destroying CO, which increases the amount of carbon atoms in the gas phase.}

               
\titlerunning{Deuteration of c-C$_3$H$_2$ towards the pre-stellar core L1544}
\authorrunning{Giers et al.}

   \maketitle
%

\section{Introduction}

Pre-stellar cores represent the initial conditions of low-mass star formation. The self-gravitating cores \citep{WardThompson1999} are dense ($n>10^5$\,cm$^{-3}$) and cold ($T<10\,K$) towards their centre \citep{Crapsi2007,Caselli2012},
showing signs of contraction motion and chemical evolution \citep{Crapsi2005}. Therefore, pre-stellar cores are ideal laboratories to study the physical and chemical processes taking place in the early stages of low-mass star formation.

In the cold and dense conditions of pre-stellar cores, the deuteration of molecules is a favoured chemical process \citep{Ceccarelli2014}, increasing the abundance of deuterated species. The deuterium fractionation is driven by the gas phase reaction (only true with respect to para states of reactants and products; \citealt{Pagani1992}):
\begin{equation}\label{equ:deuteration}
    \mathrm{H}_3^+ + \mathrm{HD} \rightleftharpoons \mathrm{H}_2\mathrm{D}^+ + \mathrm{H}_2 + 232\,\mathrm{K}.
\end{equation}
This reaction is exothermic. As the temperature decreases, the abundance of H$_2$D$^+$ ions is enhanced, which drives the deuteration of other molecules. 
As the reaction goes more efficiently from left to right at low temperatures, H$_2$D$^+$ will no longer be destroyed when meeting an H$_2$ molecule, but mostly destroyed by CO.
However, in cold and dense regions like pre-stellar cores, CO is highly depleted from the gas phase and frozen out onto dust grains \citep{Willacy1998,Caselli1999,Bacmann2003,Crapsi2005}, which further increases the deuterium fractionation \citep{DalgarnoLepp1984}.
The para-H$_2$D$^+$ is further deuterated by successive reactions with HD, leading to noticeable densities of both D$_2$H$^+$ and D$_3^+$ \citep{HilyBlant2018,Caselli2019}. All three deuterium substituted H$_3^+$ ions are driving the deuteration of other molecules, by transfer of a D$^+$ atom.
Furthermore, the level of deuteration also depends on the ortho-to-para ratio of H$_2$. The higher the ratio, the more H$_2$D$^+$ is destroyed by ortho-H$_2$, as the energy of the ground state of ortho-H$_2$ compared to para-H$_2$, $\sim170$\,K, is close to the exothermicity of reaction (\ref{equ:deuteration}).
However, the ortho-to-para ratio of H$_2$ in pre-stellar cores is usually $\sim$10$^{-3}$ \citep{Kong2015}, further promoting the deuterium fractionation.
Deuterated molecules hence have proven to be very good tracers of the central parts of dense cores that are on the verge of star formation \citep{Crapsi2007,Redaelli2019}, as a large deuterium enrichment is characteristic for these high density nuclei \citep{Crapsi2005}. 
The deuterium fraction is measured by the ratio $R_{\rm D/H}$ of the molecular tracer, dividing the column density of the D-bearing molecule by the column density of the hydrogenated isotopologue.
In addition, the level of deuteration is sensitive to the evolutionary stage of starless cores \citep{Crapsi2005,Chantzos2018}. Therefore, it can work as a chemical clock and can be used as a diagnostic tool.

Cyclopropenylidene (c-C$_3$H$_2$) is one of the most abundant and widespread molecules in the Galaxy. 
Its first detection in space was made by \cite{Thaddeus1981}, which was later confirmed by the first laboratory detection by \cite{Thaddeus1985}.
Since then, c-C$_3$H$_2$ has been observed in diffuse gas, cold dark clouds, giant molecular clouds, photodissociation regions, circumstellar envelopes, planetary nebulae, circumstellar disks, and low-mass protostars \citep{Thaddeus1985,MatthewsIrvine1985,SeaquistBell1986,Vrtilek1987,Cox1987,Madden1989,LucasLiszt2000,Teyssier2005,Qi2013,Majumdar2017}.
As the formation of c-C$_3$H$_2$ is believed to happen solely in the gas phase \citep{Park2006}, it preferentially traces dense and chemically young gas, which is gas where the C atoms have not yet been mainly locked into CO \citep{Spezzano2016b}.
However, it also suffers from depletion, and therefore it does not trace the central regions of pre-stellar cores but more external layers \citep{Spezzano2017}.
Following the detection of its singly and doubly deuterated isotopologues \citep{Gerin1987,Spezzano2013}, c-C$_3$H$_2$ is a unique probe to study the deuteration processes that take place in the gas phase in pre-stellar cores, as it is only efficiently deuterated there \citep{Spezzano2013}.

The low-mass pre-stellar core L1544 is located in the Taurus Molecular Cloud complex, one of the nearest star forming regions, and has a distance of 170\,pc \citep{Galli2019}.
The dense core is a well-known object; numerous and extensive studies have led to a profound knowledge of its physical and chemical structure \citep{Crapsi2007,Keto2004,Keto2010,Keto2015,Caselli2002a}. It is centrally concentrated \citep{WardThompson1999}, with a high central density of  around 10$^6$\,cm$^{-3}$ within the central 2000\,AU \citep{Crapsi2007,Caselli2019}, and very cold, with a temperature ranging from 12\,K in the outskirts down to $\sim$\,6\,K towards the centre \citep{Crapsi2007}.
The core has an elongated shape and shows signs of contraction motion, suggesting that it is on the verge of gravitational collapse \citep{Williams1999,Ohashi1999,Lee2001,Caselli2002a,Caselli2012}. 
Towards its centre, L1544 exhibits a high degree of CO freeze-out (93\%; \citealt{Caselli2002a,Crapsi2005}), and a high level of deuteration \citep{Crapsi2005,Redaelli2019}. 
It is chemically rich, showing spatial inhomogeneities in the distribution of molecular emission \citep{Spezzano2017,Redaelli2019,ChaconTanarro2019}. 

In this work we present emission maps of c-C$_3$H$_2$, the singly and doubly deuterated isotopologues, c-C$_3$HD and c-C$_3$D$_2$, as well as the isotopologue c-H$^{13}$CC$_2$H towards L1544.
In Sect. \ref{observations}, we describe the observations, followed by the results in Sect. \ref{results}. 
The analysis in Sect. \ref{analysis} is divided into the derivation of the c-C$_3$H$_2$ deuteration maps and a comparison of the dust emission peak column density with non-LTE simulations.
We discuss the results in Sect. \ref{discussion} and present our conclusions in Sect. \ref{conclusion}.

\section{Observations}\label{observations}

The emission maps presented in this paper were obtained using the IRAM 30\,m single dish telescope located in Pico Veleta, Spain, in October 2013. The observed $2.5'\times2.5'$ on-the-fly (OTF) maps are centred on the source dust emission peak ($\alpha_{2000}=05^\mathrm{h}04^\mathrm{m}17^\mathrm{s}.21$, $\delta_{2000}=+25^\circ10'42''.8$, \citealt{WardThompson1999}). We applied position switching, with the reference position set at (-180$''$,180$''$) offset with respect to the map centre. The EMIR E090 receiver was used along with the Fourier transform spectrometer backend (FTS), with a spectral resolution of 50\,kHz. 
The antenna moved along an orthogonal pattern of linear paths separated by $8''$ intervals, corresponding roughly to one third of the beam full width at half maximum (FWHM). The weather conditions during the mapping were good ($\tau\sim0.03$), with a typical system temperature of $T_\mathrm{sys}\sim$\,90$-$100\,K.

The observed transitions are summarised in Table \ref{LineParam}.
Figure~\ref{FigIntInt} shows the corresponding integrated intensity maps for each molecule. 
The data processing was done using the GILDAS software \citep{Pety2005} and Python. All emission maps presented in this paper have been gridded to a pixel size of $8''$ with the CLASS software in the GILDAS packages; this corresponds to 1/3\,-1/4 of the actual beam size, depending on the frequency.
The antenna temperature $T_A^*$ was converted to the main beam temperature $T_\mathrm{mb}$ using the relation $T_\mathrm{mb}=F_\mathrm{eff}/B_\mathrm{eff}\cdot T_A^*$. The corresponding values for the 30\,m forward ($F_\mathrm{eff}$) and main beam efficiency ($B_\mathrm{eff}$) ratios are given in Table \ref{LineParam}.

\begin{table*}
    \caption[]{Observed lines and observation details.}
    \label{LineParam}
    $$
    \begin{tabular}{llrcccccc}
    \hline\hline
    \noalign{\smallskip}
     Molecule & Transition & Frequency\,\tablefootmark{a} & rms & HPBW & $F_\mathrm{eff}/B_\mathrm{eff}$ & Velocity range\,\tablefootmark{b} & Reference \\
     & $(J_{K'_aK'_c}'-J_{K_aK_c})$ & (GHz) & (mK)  & (") &  & (km\,s$^{-1}$) & \\
     \noalign{\smallskip}
     \hline
     \noalign{\smallskip}
     c-C$_3$H$_2$ & $3_{22} - 3_{13}$ & 84.728 & \textbf{22} & 30.6 & 0.95/0.838 & 6.90$-$7.42 & 1 \\
     c-H$^{13}$CC$_2$H & $2_{12} - 1_{01}$ & 84.186 & \textbf{22} & 30.8 & 0.95/0.839 & 6.83$-$7.35  & 2 \\
     c-C$_3$HD & $3_{03} - 2_{12}$ & 104.187 & \textbf{27} & 24.9 & 0.95/0.802  & 6.78$-$7.48 & 2\\
     c-C$_3$D$_2$ & $3_{13} - 2_{02}$ & 97.762 & \textbf{21} & 26.5 & 0.95/0.814 & 6.89$-$7.49 & 3 \\
     \noalign{\smallskip}
    \hline 
    \end{tabular}
    $$
    \tablefoot{
    \tablefoottext{a}{Extracted from Cologne database for molecular spectroscopy \citep{Mueller2001}}
    \tablefoottext{b}{Velocity ranges where the integrated emission has been computed.}
    }
    \tablebib{
    (1) \cite{Thaddeus1985}; (2) \cite{Bogey1987}; (3) \cite{Spezzano2012}.
    }
    
\end{table*}

%
   
\section{Results}\label{results}

The integrated intensities were derived calculating the zeroth moment of the emission maps, integrating over the velocity ranges given in Table \ref{LineParam}. 
The ranges  contain all velocity channels whose map showed a signal-to-noise ratio larger than or equal to 3.
To enable the comparison between the molecules, all maps are convolved to an angular resolution of $31''$, which corresponds to the half power beam width (HPBW) of the largest observed beam (c-H$^{13}$CC$_2$H$_2$).

\begin{figure*}
   \centering
   \includegraphics[width=.49\textwidth]{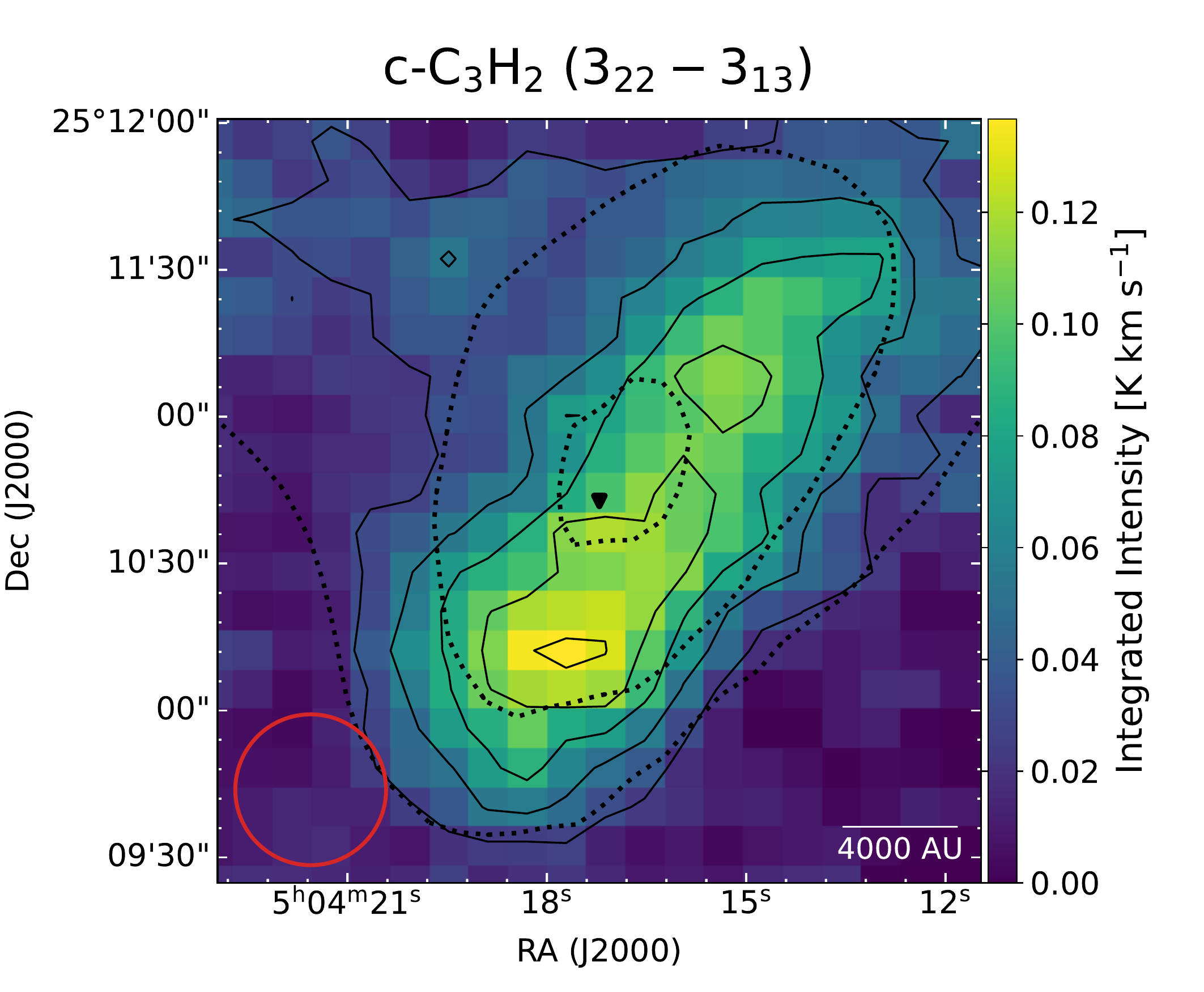}
    \includegraphics[width=.49\textwidth]{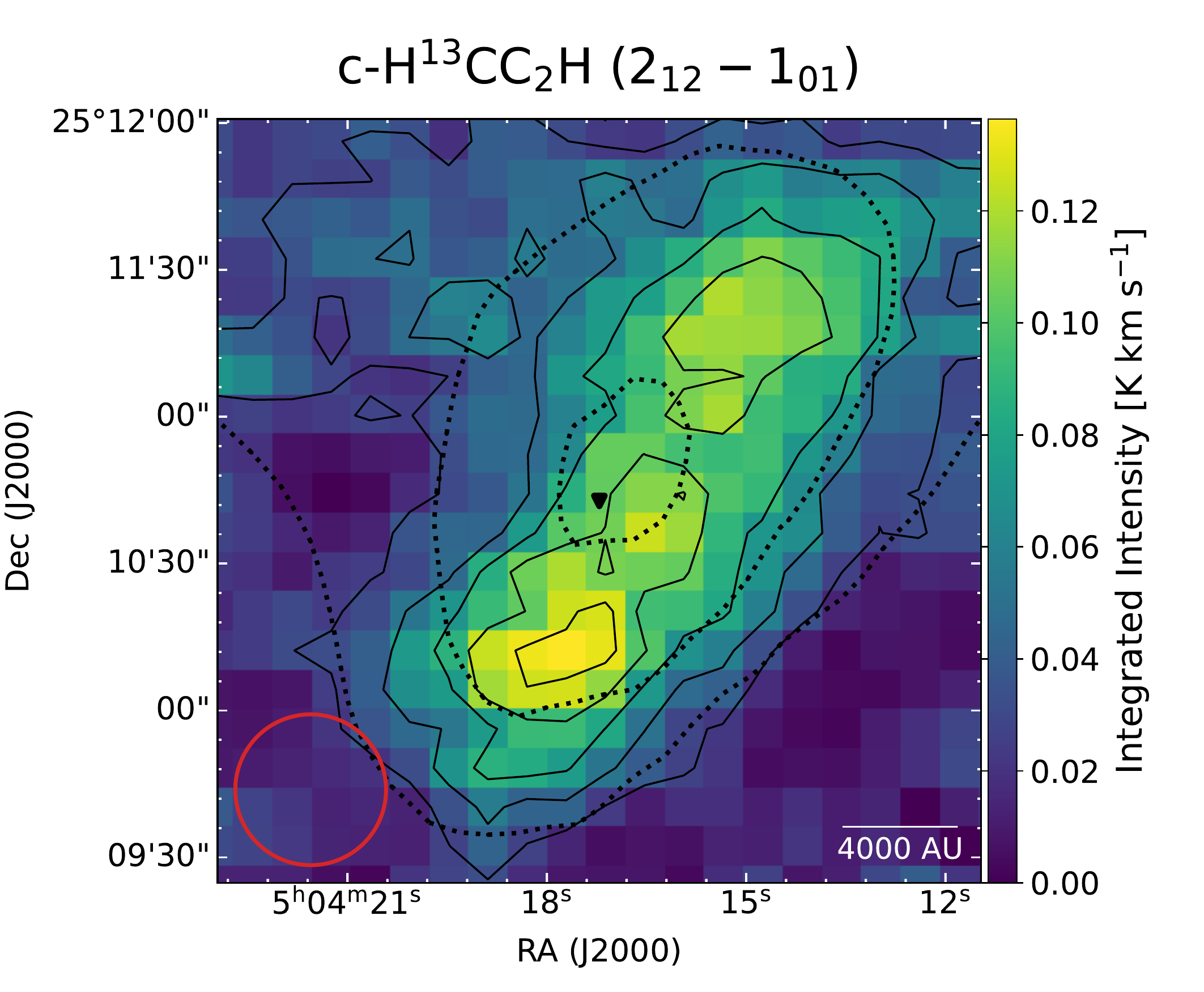}
    \includegraphics[width=.49\textwidth]{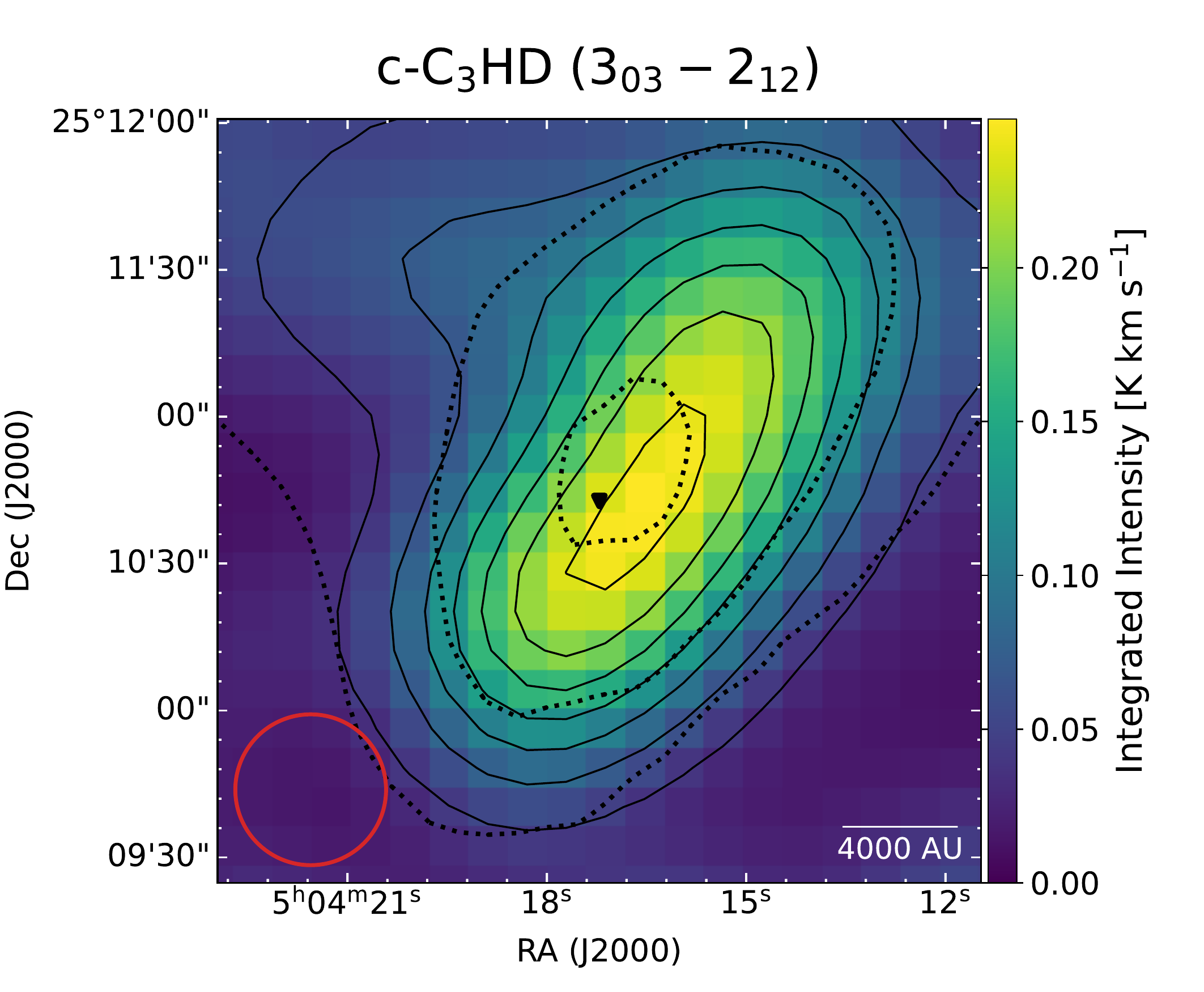}
    \includegraphics[width=.49\textwidth]{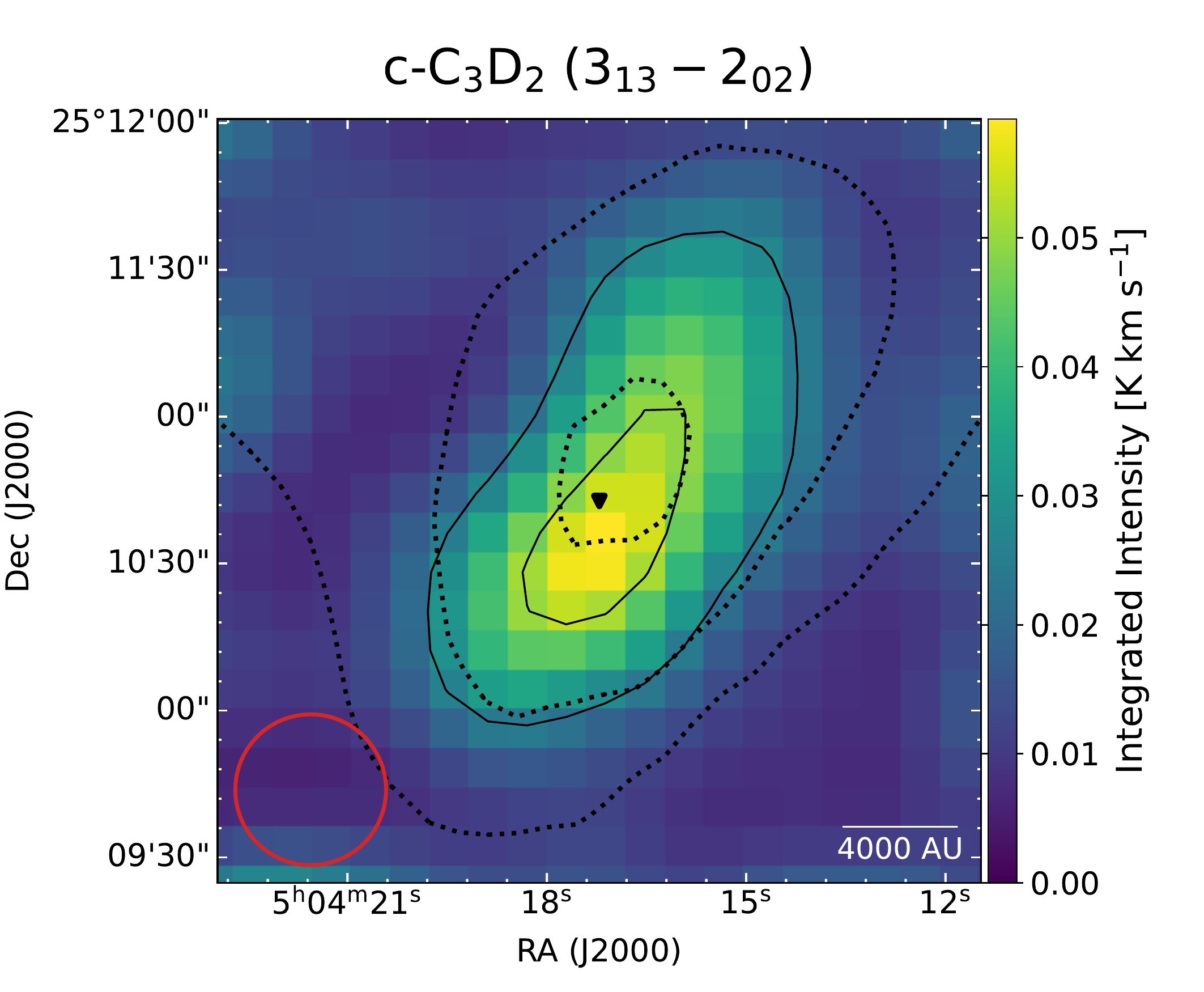}
   \caption{Integrated intensity maps of the observed transitions. The black dashed lines represent the 90\%, 50\% and 30\% of the H$_2$ column density peak value derived from \textit{Herschel} maps \citep{Spezzano2016b}, 2.8$\times$10$^{22}$\,cm$^{-2}$. The solid lines represent contours of the integrated intensity starting at $3\sigma$ with steps of $3\sigma$. From top to bottom and left to right, the average errors on the integrated intensity are 9, 9, 12, 8, in units of mK\,km\,s$^{-1}$. The dust peak is indicated by the black triangle. The beam size of the 30\,m telescope, $\rm HPBW=31''$, is shown by the red circle in the bottom left of each panel, and the scalebar is shown in the bottom right corners.}
              \label{FigIntInt}
    \end{figure*}

\paragraph{c-C$_3$H$_2$.} 
The integrated intensity map of para-c-C$_3$H$_2$ ($3_{22}-3_{13}$) is shown in the top left panel of Fig.~\ref{FigIntInt}. The emission peak is located to the south-east of the dust peak. This is driven by a non-uniform illumination from the interstellar radiation field, which dominates towards the southern part of the core and leads to more C atoms in the gas phase and subsequently an increased formation of c-C$_3$H$_2$, as discussed in \cite{Spezzano2016b}.

\paragraph{c-H$^{13}$CC$_2$H$_2$.}
The distribution of the c-H$^{13}$CC$_2$H$_2$ integrated emission (Fig.~\ref{FigIntInt}, top right panel) follows the distribution of c-C$_3$H$_2$, with an emission peak located at the c-C$_3$H$_2$ peak and similar values of integrated intensity. 

\paragraph{c-C$_3$HD.}
The integrated intensity map of c-C$_3$HD, bottom left panel of Fig.~\ref{FigIntInt}, peaks in an elongated region passing through the dust peak, where temperature and density are suitable for the deuteration process to take place. 
The emission peak follows the direction of the major axis and extends more to the north-west than to the south-east, suggesting a flattened central structure, predicted by simulations of magnetized core contraction (see e.g. \citealt{Caselli2019}). 

\paragraph{c-C$_3$D$_2$.}
The integrated intensity map of c-C$_3$D$_2$ is shown in the bottom right panel of Fig.~\ref{FigIntInt}. The emission peak is located very close to the dust peak, confirming that closer to the centre of the core the level of deuteration is higher. In addition, the emission is less extended than for c-C$_3$H$_2$ and c-C$_3$HD, probably tracing a higher density and more central shell of the core. 
 
\section{Analysis}\label{analysis}

\subsection{Column densities}\label{columndensity}
In this work, all column densities are calculated using the assumption of optically thin emission, as presented in \cite{Mangum2015}. Furthermore, we apply the approximation of a constant excitation temperature throughout the core (CTex), following \cite{Caselli2002b} and \cite{Redaelli2019}:
\begin{equation}\label{EquColDens}
    N=\frac{8\pi\nu^3}{c^3}\frac{Q_\mathrm{rot}(T_\mathrm{ex})}{g_uA_\mathrm{ul}}\left[J_\nu(T_\mathrm{ex})-J_\nu(T_\mathrm{bg})\right]^{-1}\frac{\mathrm{e}^{\frac{E_u}{kT_\mathrm{ex}}}}{\mathrm{e}^{\frac{h\nu}{kT_\mathrm{ex}}}-1}\int T_\mathrm{mb}\mathrm{d}v\;,
\end{equation}
where $Q_\mathrm{rot}(T_\mathrm{ex})$ is the partition function of the molecule at an excitation temperature $T_\mathrm{ex}$, $g_u$ and $E_u$ are the degeneracy and energy of the upper level of the transition, respectively, $A_\mathrm{ul}$ the Einstein coefficient for spontaneous emission, $T_\mathrm{bg}=2.73$\,K the temperature of the cosmic mircowave background, $J(T)$ the Rayleigh-Jeans equivalent temperature and $T_\mathrm{mb}$ the main beam temperature. The corresponding parameters for each transition used in the derivation of the column density are listed in Table \ref{TabNcolParam}.

As excitation temperature, we use 6\,K for the main species and for the $^{13}$C-bearing isotopologue (see \citealt{Gerin1987}). For the deuterated isotopologues, we choose an excitation temperature of 5\,K, following \cite{Spezzano2013}.
The effect of the excitation temperature on the derived column density ratios was found to be small, with a change of few percent upon a variation of $\pm$\,1\,K (as stated in \citealt{Spezzano2013}).\\
Our assumption of a constant excitation temperature applied across the maps follows the analysis presented in the appendix of \cite{Redaelli2019}. There, the authors prove that their results are not sensitive to the variations of the excitation temperature across the core and thus the analysis can be done assuming a constant value. As our observations have similar sized maps and angular resolution, we apply this assumption.

The derived column density maps are shown in Fig.~\ref{FigNcolmaps} in the Appendix.
The morphologies of the maps follow the integrated intensity maps shown in Fig.~\ref{FigIntInt}. 
The main species shows the highest column densities, peaking at 2.0$\times$10$^{13}$\,cm$^{-2}$. The $^{13}$C isotopologue and the singly deuterated form have a lower peak column density than the main isotopologue, with peak values around 1.3$\times$10$^{12}$\,cm$^{-2}$ and 4.0$\times$10$^{12}$\,cm$^{-2}$. 
Furthermore, the doubly deuterated isotopologue only peaks at $\sim$\,6.3$\times$10$^{11}$\,cm$^{-2}$, also one order of magnitude less in column density than its precursor molecule, c-C$_3$HD. On average, the uncertainties on the column density maps are 1$\times10^{12}$\,cm$^{-2}$, 9$\times10^{10}$\,cm$^{-2}$, 2$\times10^{11}$\,cm$^{-2}$ and 1$\times10^{11}$\,cm$^{-2}$ for c-C$_3$H$_2$, c-H$^{13}$CC$_2$H$_2$, c-C$_3$HD and c-C$_3$D$_2$, respectively. The error maps are derived through standard error propagation from the rms maps, which are calculated from the velocity channel maps that do not show any signal.

Furthermore, we extracted spectra at the dust peak and the c-C$_3$H$_2$ peak of L1544, respectively, which are shown in Fig.~\ref{FigAveSpectra} in Appendix \ref{section:averagedspectra}.
Using Gaussian fitting in Python, we derive the column densities at the two peaks. The resulting best-fit parameters along with the corresponding column densities and optical depths are summarised in Table \ref{tab:peakfitobs}.
In general, the derived column densities of the deuterated isotopologues are higher towards the dust peak than the c-C$_3$H$_2$ peak. The derived optical depths show that all observed lines are optically thin.
However, by multiplying the column density of the $^{13}$C isotopologue with the isotopic ratio for the local interstellar medium, $^{12}$C/$^{13}$C\,=\,68 \citep{Milam2005}, divided by two (to account for the degeneracy of c-H$^{13}$CC$_2$H regarding the position of the $^{13}$C atom), we obtain an additional way to determine the column density of the main isotopologue. 
Despite the large uncertainty of the resulting value, $N(^{13}$C,CTex$)=(44\pm10)\times10^{12}$\,cm$^{-2}$, it might suggest that the c-C$_3$H$_2$ ($3_{33}-3_{13})$ transition is moderately optically thick.

\begin{table*}
    \caption[]{Parameters used in the derivation of the column densities.}
    \label{TabNcolParam}
    $$
    \begin{tabular}{lccrcr}
        \hline
        \hline
        \noalign{\smallskip}
        Molecule & T$_\mathrm{ex}$ (K) & Q(T$_\mathrm{ex})$ & E$_u$ (K) & A $(10^{-5}$s$^{-1})$ & g$_u$ \\
        \noalign{\smallskip}
        \hline
        \noalign{\smallskip}
        c-C$_3$H$_2$      ($3_{22}-3_{13}$) & 6 & 37.68 & 16.14 & 1.04 & 7  \\
        c-H$^{13}$CC$_2$H ($2_{12}-1_{01}$) & 6 & 38.69 & 6.33  & 2.17 & 10 \\
        c-C$_3$HD         ($3_{03}-2_{12}$) & 5 & 50.10 & 10.86 & 3.96 & 21 \\
        c-C$_3$D$_2$      ($3_{13}-2_{02}$) & 5 & 86.37 & 9.89  & 3.88 & 42 \\
        \noalign{\smallskip}
        \hline
        \noalign{\smallskip}
    \end{tabular}
    $$
\end{table*}

\begin{table*}[h]
    \centering
    \caption{Properties of the spectral lines extracted at the dust peak and the c-C$_3$H$_2$ peak.}
    \begin{tabular}{l c c c l l l c}
    \hline\hline 
    \noalign{\smallskip}
     Molecule   & $T_\mathrm{mb}$ & $\tau$\,\tablefootmark{a} & rms & $W$ & $\rm{v}_\mathrm{LSR}$ & $\Delta \rm{v}$\,\tablefootmark{b} & $N$\,\tablefootmark{a} \\
      & (K) &  & (mK) & (K\,km\,s$^{-1}$) & (km\,s$^{-1}$) & (km\,s$^{-1}$) & (10$^{12}$\,cm$^{-2}$) \\
      \noalign{\smallskip}
      \hline
      \noalign{\smallskip}
      \multicolumn{8}{c}{\textbf{Dust peak}} \\
     c-C$_3$H$_2$      & 0.28(4) & 0.10(1) & 35 & 0.10(2)  & 7.20(2) & 0.33(5) & 17(3)\\
     c-H$^{13}$CC$_2$H & 0.19(3) & 0.07(1) & 32 & 0.12(3)  & 7.15(4) & 0.6(1)  & 1.3(3)\\
     c-C$_3$HD         & 0.59(3) & 0.36(2) & 39 & 0.27(2)  & 7.20(1) & 0.43(3) & 5.3(4)\\
     c-C$_3$D$_2$      & 0.14(2) & 0.07(1) & 27 & 0.07(2)  & 7.19(4) & 0.47(9) & 0.88(3)\\
     \noalign{\smallskip}
     \hline 
     \noalign{\smallskip}
     \multicolumn{8}{c}{\textbf{c-C$_3$H$_2$ peak}} \\
     c-C$_3$H$_2$      & 0.49(4) & 0.18(1) & 28 & 0.14(2) & 7.21(1) & 0.27(3) & 23(3)\\
     c-H$^{13}$CC$_2$H & 0.32(3) & 0.10(1) & 32 & 0.13(2) & 7.21(2) & 0.40(4) & 1.4(2)\\
     c-C$_3$HD         & 0.74(4) & 0.47(2) & 39 & 0.24(2) & 7.29(1) & 0.31(2) & 4.7(4)\\
     c-C$_3$D$_2$      & 0.16(3) & 0.08(2) & 27 & 0.04(1) & 7.30(2) & 0.25(6) & 0.5(1)\\
     \noalign{\smallskip}
     \hline
     \noalign{\smallskip}
    \end{tabular}
    \tablefoot{\tablefoottext{a}{Derived using the excitation temperatures listed in Table~\ref{TabNcolParam}.}\tablefoottext{b}{Observed linewidth.}} 
    \label{tab:peakfitobs}
\end{table*}

\subsection{Deuterium fraction}\label{deuteriumfraction}
The deuterium fraction maps are derived by dividing the column density maps of the deuterated isotopologues by that of the main species, pixel by pixel.
To avoid issues with the optical depth of the main isotopologue, we use the column density map for c-C$_3$H$_2$ derived from the $^{13}$C isotopologue multiplied by the isotopic ratio for the local interstellar medium divided by two, $^{12}\mathrm{C}/^{13}\mathrm{C}$\,=\,68/2\,=\,34. 
\cite{Colzi2020} showed that in the local interstellar medium the ratio is molecule-dependent and varies with time, volume density, and temperature, and can deviate significantly from 68. For example, for a fixed density and temperature, it may change by a factor of 2 between 10$^5$ and 10$^6$ years of cloud evolution, depending on the molecule.
It is important to note that the emission of the different isotopologues of c-C$_3$H$_2$ across the core comes mostly from a layer of the core with volume density similar to the critical density of the transitions (a few $\times$10$^5$\,cm$^{-3}$).
Due to this, we only expect minor variation of the $^{12}$C/$^{13}$C ratio of c-C$_3$H$_2$ across L1544. 
While variations of the $^{12}$C/$^{13}$C ratio with respect to the ISM value will affect the absolute values of the column densities and column density ratios we derive, they will not affect the overall morphology of the maps and the conclusions that we draw here.  

\begin{figure*}
    \includegraphics[width=.99\textwidth]{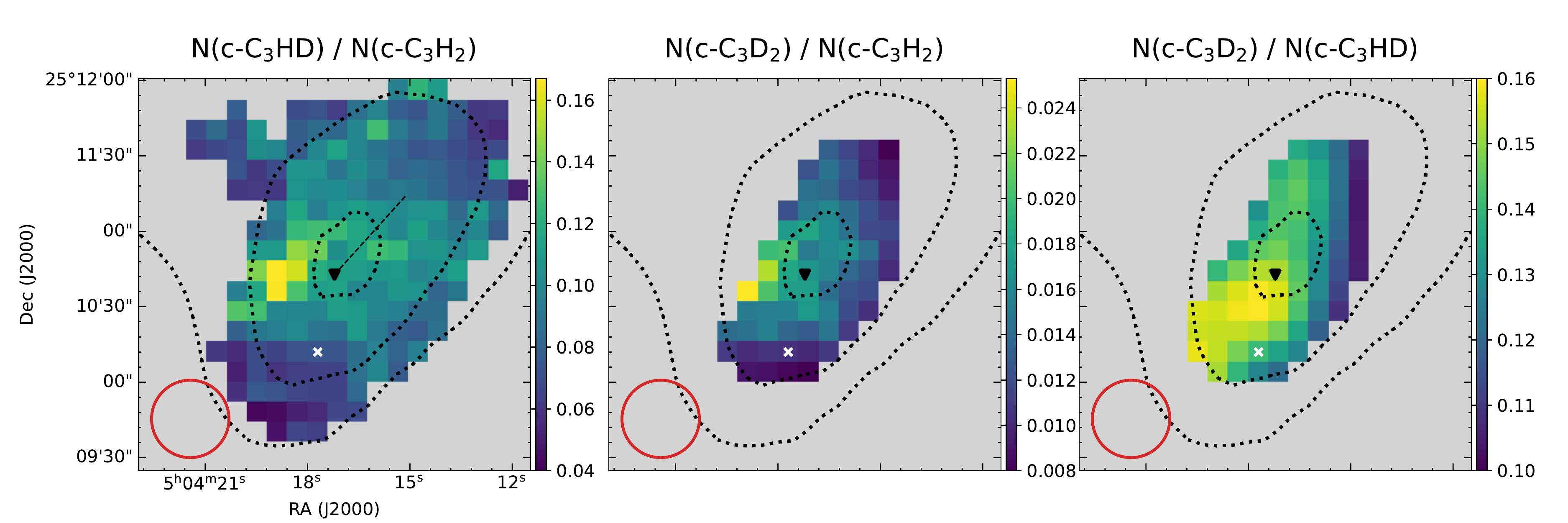}
    \caption{Deuteration maps of c-C$_3$H$_2$. Only pixels above $3\sigma$ are plotted, pixels below are grey. The black dashed lines represent 90\%, 50\% and 30\% of the H$_2$ column density peak value derived from Herschel maps \citep{Spezzano2016b}, 2.8$\times$10$^{22}$\,cm$^{-2}$. The black triangle indicates the dust continuum peak \citep{WardThompson1999}, the white cross shows the c-C$_3$H$_2$ peak. The HPBW is indicated by a red circle in each panel. The dashed line in the left panel shows the cut used to produce Figure~\ref{Figdhcut}.}
    \label{FigDeutFrac}%
\end{figure*}

The resulting deuterium fraction maps are shown in Fig.~\ref{FigDeutFrac}, where we show pixels with signal above $3\sigma$. The dust continuum peak is indicated by a black triangle, while the white cross shows the c-C$_3$H$_2$ peak.
The map of $R_{\rm D/H}$(c-C$_3$H$_2)=N($c-C$_3$HD)/$N$(c-C$_3$H$_2$)  shows a peak located $\sim$20$''$ to the east of the dust peak, reaching a maximum of $0.17\,\pm\,0.01$. 
In the map of $R_{\rm D2/H2}$(c-C$_3$H$_2)=N$(c-C$_3$D$_2$)/N(c-C$_3$H$_2$), the peak is located $\sim$20$''$ east of the dust peak, with a maximum of $0.025\,\pm\,0.003$. 
The level of deuteration in the map of $R_{\rm D2/HD}$(c-C$_3$H$_2) =N$(c-C$_3$D$_2$)/$N$(c-C$_3$HD) ranges between 0.10 and 0.16, while the peak is located at the dust peak, with a maximum of $0.16\,\pm\,0.03$. 
The distribution shows a concentration of pixels with values of 0.15\,-\,0.16 spreading between the dust peak and the c-C$_3$H$_2$ peak.

\subsection{Non-LTE modelling}\label{modeling}

We test the assumption of local thermodynamic equilibrium (LTE) used in the derivation of the column densities by modelling the spectra at the dust peak, using non-LTE simulations. For this, we run the radiative transfer codes RADEX \citep{VanDerTak2007} and MOLLIE \citep{Keto1990,Keto2004}.

The radiative transfer code RADEX is one-dimensional, using the escape probability formulation to simplify the calculations. For this work, we apply the geometry of a static, spherically symmetric and uniform medium for the escape probability used to derive the column densities.

\begin{figure}
   \centering
   \includegraphics[width=\hsize]{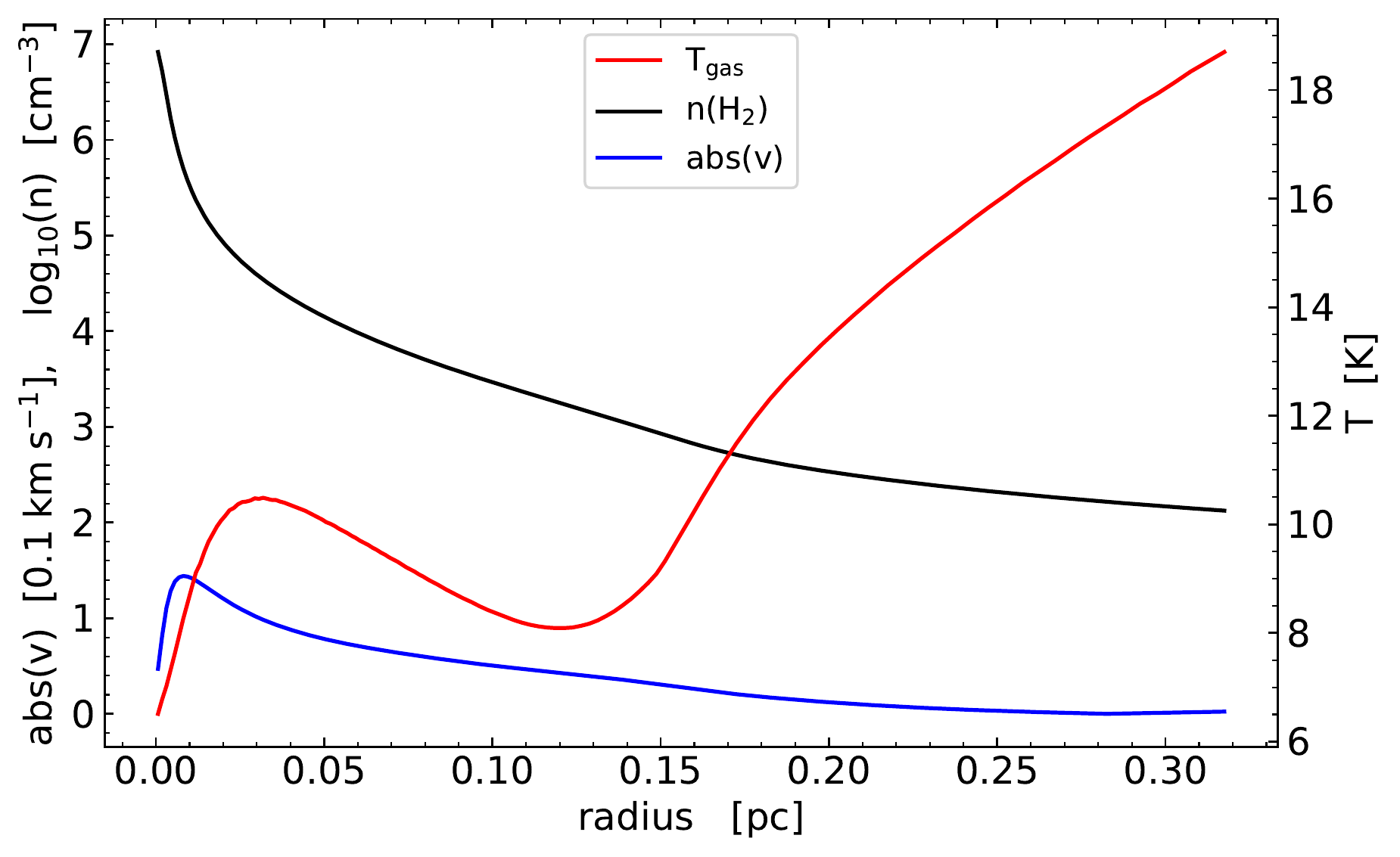}
   \caption{Profiles of the gas temperature (red), H$_2$ number density (black, in logarithmic scale) and infall velocity (blue, in units of 0.1\,km\,s$^{-1}$) for the L1544 model developed by \cite{Keto2015}. The velocity in the model is negative, but here it is shown as positive for better readability.}
              \label{FigPhysicalModel}
\end{figure}

In contrast to RADEX, MOLLIE solves the radiative transfer equation on a three-dimensional grid, using a spherically symmetric model of the source. 
This physical model is taken from \cite{Keto2015}, who describe an unstable quasi-equilibrium Bonnor-Ebert sphere with a peak central H$_2$ volume density of $n_0\approx 10^7$\,cm$^{-3}$ and a central gas temperature of 6\,K. The model provides the infall velocity, density and gas temperature structure for MOLLIE (see Fig.~\ref{FigPhysicalModel}). 
The simulated molecular abundance profiles are derived using a state-of-the-art gas-grain chemical model \citep{Sipila2019}, assuming an external visual extinction of $A_\mathrm{V}=2\,$mag to account for the fact that L1544 is embedded in a molecular cloud.
We employ the same initial abundances and chemical networks as in \cite{Sipila2019}. The initial abundances are reproduced here in Table~\ref{tab:initialabundances}. We assume `standard' values for various model parameters; we set a constant cosmic-ray ionization rate $\zeta = 1.3 \times 10^{-17} \, \rm s^{-1}$, a grain radius of 0.1\,$\mu$m (monodisperse grains), and a grain material density of $2.5 \, \rm g \, cm^{-3}$. 
\cite{Redaelli2021} found a cosmic-ray ionization rate of $\zeta = \sim 3 \times 10^{-17} \, \rm s^{-1}$, which is a factor of $\sim2$ different to the value we apply.
In Appendix~\ref{section:testchemicalmodel}, we discuss the effect of an increased cosmic-ray ionization rate on the chemistry.
The inital ortho-to-para ratio of H$_2$ is set to $10^{-3}$ and evolves with the rest of the chemistry.
We adopt a two-phase description of gas-grain chemistry, that is, the ice on the grain surfaces is treated as a single reactive layer. Molecules may be desorbed thermally, or via cosmic ray induced desorption, photodesorption, or chemical desorption (with 1\% efficiency).

With the two different codes we intend to test how a rather `simple' and often used code like RADEX, which only applies a general spherical core geometry with constant molecular abundance, volume density and temperature, performs compared to a more sophisticated but time-consuming simulation like MOLLIE, which considers the specific physical structure of a source and molecular abundance profiles across it.

\begin{table}
        \centering
        \caption{Initial abundances (with respect to $n_{\rm H} \approx 2\,n({\rm H_2})$) used in the chemical modelling.}
        \begin{tabular}{l|l}
                \hline
                \hline
                Species & Abundance\\
                \hline
                $\rm H_2$ & $5.00\times10^{-1}\,^{(a)}$\\
                $\rm He$ & $9.00\times10^{-2}$\\
                $\rm C^+$ & $1.20\times10^{-4}$\\
                $\rm N$ & $7.60\times10^{-5}$\\
                $\rm O$ & $2.56\times10^{-4}$\\
                $\rm S^+$ & $8.00\times10^{-8}$\\
                $\rm Si^+$ & $8.00\times10^{-9}$\\
                $\rm Na^+$ & $2.00\times10^{-9}$\\
                $\rm Mg^+$ & $7.00\times10^{-9}$\\
                $\rm Fe^+$ & $3.00\times10^{-9}$\\
                $\rm P^+$ & $2.00\times10^{-10}$\\
                $\rm Cl^+$ & $1.00\times10^{-9}$\\
                $\rm HD$ & $1.60\times10^{-5}$\\
                \hline
        \end{tabular}
        \label{tab:initialabundances}
        \tablefoot{$^{(a)}$ The initial $\rm H_2$ ortho/para ratio is $1 \times 10^{-3}$.}
\end{table}

\subsubsection{RADEX: Results}
The results are obtained by manually fitting the synthetic spectra derived by RADEX to the observed spectra at the dust peak given in Figure \ref{FigAveSpectra} in the appendix. This was done using the CASSIS software \citep{VastelCASSIS2015}, assuming $n(\mathrm{H}_2)=10^6$\,cm$^{-3}$. We set $T_\mathrm{kin}=5$\,K for c-C$_3$D$_2$ ($3_{13}-2_{02}$) and $T_\mathrm{kin}=8$\,K for c-C$_3$H$_2$ ($3_{22}-3_{13}$) and c-C$_3$HD ($3_{03}-2_{12}$), corresponding to the lowest temperatures with available collision rates.
The collision rates of c-C$_3$H$_2$, c-C$_3$HD and c-C$_3$D$_2$ with H$_2$ were provided by \cite{BenKhalifa2019} for the main species and Ben Khalifa et al. (in prep.) for the isotopologues. 
Because of its different symmetry, the c-H$^{13}$CC$_2$H isotopologue collisional rates cannot be deduced from the other rates, and they have not been computed. Hence, we do not model the $^{13}$C isotopologue in the non-LTE formalisms.

The total column densities corresponding to the best fit solutions of RADEX are listed in column two of Table \ref{tab:compareNcol}. The corresponding uncertainties of RADEX are estimated by varying the column density to result in a change of the peak temperature by 25\%. To derive the total column densities of c-C$_3$H$_2$ and  c-C$_3$D$_2$, the RADEX output column densities are multiplied by the ortho-to-para ratio, which is 3:1 and 1:2 for the two molecules, respectively. This is done because the collision rates used in the calculations treat ortho and para as two different species, so they are modelled separately. 

Using RADEX, we can, within the errorbars, reproduce the column density of c-C$_3$H$_2$ observed at the dust peak. The value for c-C$_3$HD is only slightly underestimated; c-C$_3$D$_2$, however, is overestimated by a factor of $\sim2$.

\subsubsection{MOLLIE: Results}
As output, MOLLIE produces a synthetic spectral line profile. The molecular column density is obtained by integrating the abundance profile multiplied by the gas density, and convolved to a desired beam size.
The collisional coefficients needed for the line radiative transfer are provided by \cite{BenKhalifa2019} and Ben Khalifa et al. (in prep).

The total column densities corresponding to the best fit solutions of MOLLIE are listed in columns three and four of Table \ref{tab:compareNcol}. The values for c-C$_3$H$_2$ and  c-C$_3$D$_2$ are derived by multiplying the MOLLIE output column densities by the ortho-to-para ratio (3:1 and 1:2 for c-C$_3$H$_2$ and c-C$_3$D$_2$, respectively) to account for the separate treatment of ortho and para species during the modelling. The errors of the column densities derived with MOLLIE lie roughly within a factor of 2, and are constrained by the resolution of the time steps of the chemical model.

\begin{figure}
   \centering
   \includegraphics[width=\hsize]{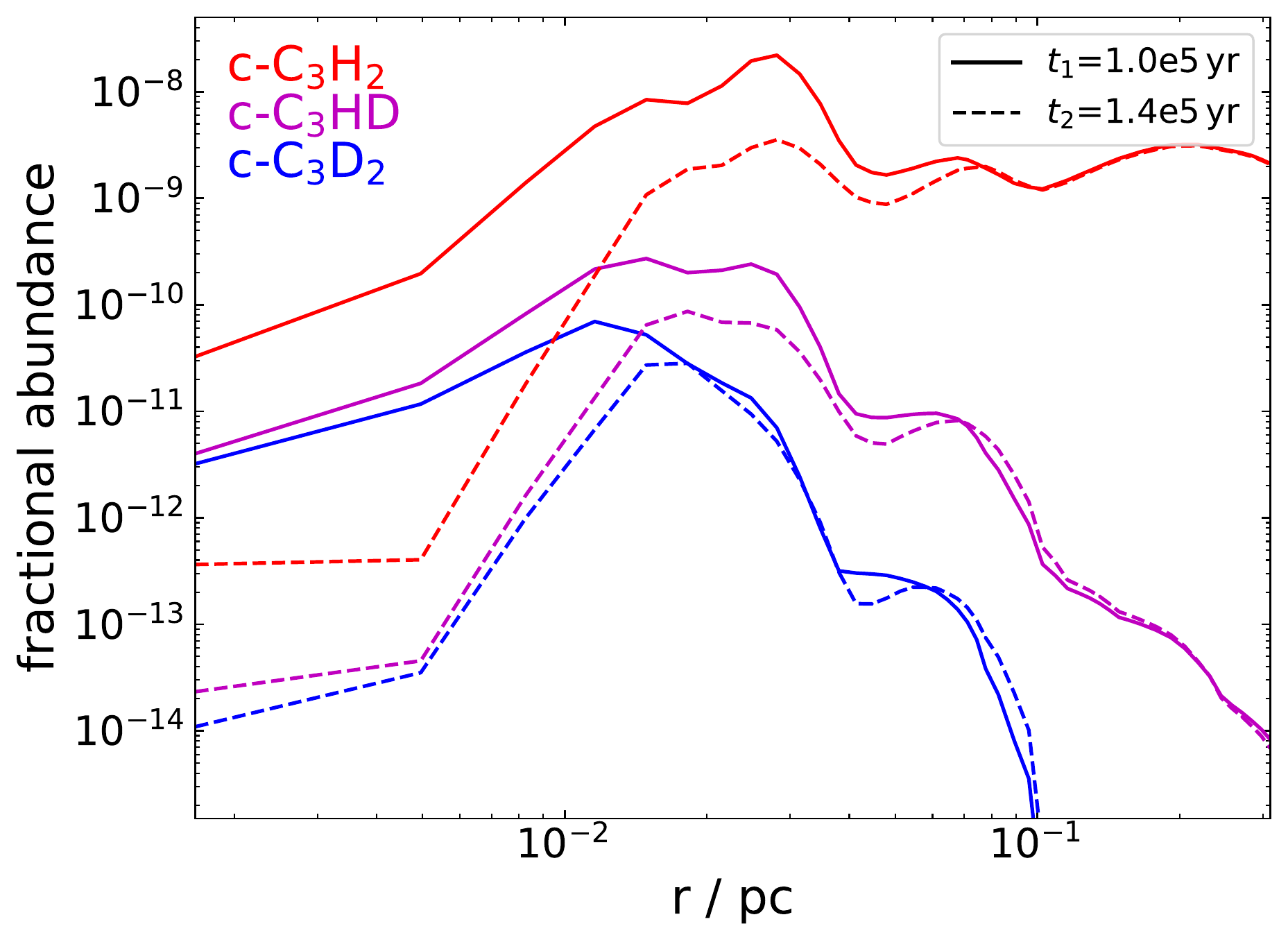}
   \caption{Total molecular fractional abundance profiles with respect to H$_2$ used to obtain the MOLLIE best fit solutions. Plotted are c-C$_3$H$_2$ (red, c-C$_3$HD (pink) and c-C$_3$D$_2$ (blue), at two different time steps of the chemical model, $t_1$\,=\,1.0$\times$10$^5$\,yrs (solid lines) and $t_2$\,=\,1.4$\times$10$^5$\,yrs (dashed lines).}
              \label{Fig6Abundances}
\end{figure}

Figure \ref{Fig6Abundances} shows the total molecular abundance profiles of c-C$_3$H$_2$, c-C$_3$HD and c-C$_3$D$_2$ with respect to H$_2$ at the two best fitting time steps of the chemical model, $t_1$\,=\,1.0$\times$10$^5$\,yrs (solid lines) and $t_2$\,=\,1.4$\times$10$^5$\,yrs (dashed lines). 
At the first time step, c-C$_3$H$_2$ peaks at $\sim$\,5$\times$10$^3$\,AU, while c-C$_3$HD and c-C$_3$D$_2$ peak at $\sim$\,3$\times$10$^3$\,AU, slightly shifted towards the centre of the core. At the second time step, all peaks are shifted to larger radii, as the molecules get depleted in the centre of the core with the chemistry evolving. Furthermore,  c-C$_3$H$_2$ shows a rather flat profile in the outer radii, being abundant also in less dense parts, while the D-bearing species quickly drop off after they peak, staying more concentrated within the central parts.

We only vary the time step in the chemical model in our analysis but not any other input parameters of the model.
Therefore, we tested the effect of the various parameters by running four single-point simulations. 
The results of these test simulations are presented in Appendix~\ref{section:testchemicalmodel}.

Figure \ref{FigMollieResults} compares the spectra modelled with MOLLIE (coloured) at the two time steps $t_1$\,=\,1.0$\times$10$^5$\,yrs (solid lines) and $t_2$\,=\,1.4$\times$10$^5$\,yrs (dashed lines) with the observed spectra (black). The later time step shows a very good fit with the observed spectrum of c-C$_3$H$_2$. The deuterated isotopologues, on the other hand, are both underestimated by a factor of 3. They are better reproduced by $t_1$, where the main species is overestimated by a factor of 7. 
The same applies for the derived column densities; at $t_2$, the observed values of c-C$_3$HD and c-C$_3$D$_2$ are reproduced, while, within the error bars, c-C$_3$H$_2$ is reproduced at $t_1$. 
One would expect the main species to match with the earlier time step, as the deuteration process 
is expected to happen late in the core evolution,
with the deuterated isotopologues peaking at a later time. 
However, it is important to note that the chemical code used to produce the abundance profiles is static (i.e., the physical structure does not evolve with time).  
Thus, the timesteps of the chemical model are not linked to the dynamical evolution of the cloud. 
This behaviour can be explained by the fact that the chemical time scale is approximately inversely proportional to the volume density (more reactants imply faster rates), so the central regions have faster chemical time scales compared to the outer less dense regions. When using a static core, this could lead to inconsistencies, as, in reality, a contracting core spends longer times at low densities, while the denser regions form only at a later stage of its evolution.
Therefore, the small difference between the time steps and the large difference between the resulting spectra demonstrates the sensitivity of the chemical model to the chosen time step and points out its limitations.

\begin{figure*}
   \includegraphics[width=.99\textwidth]{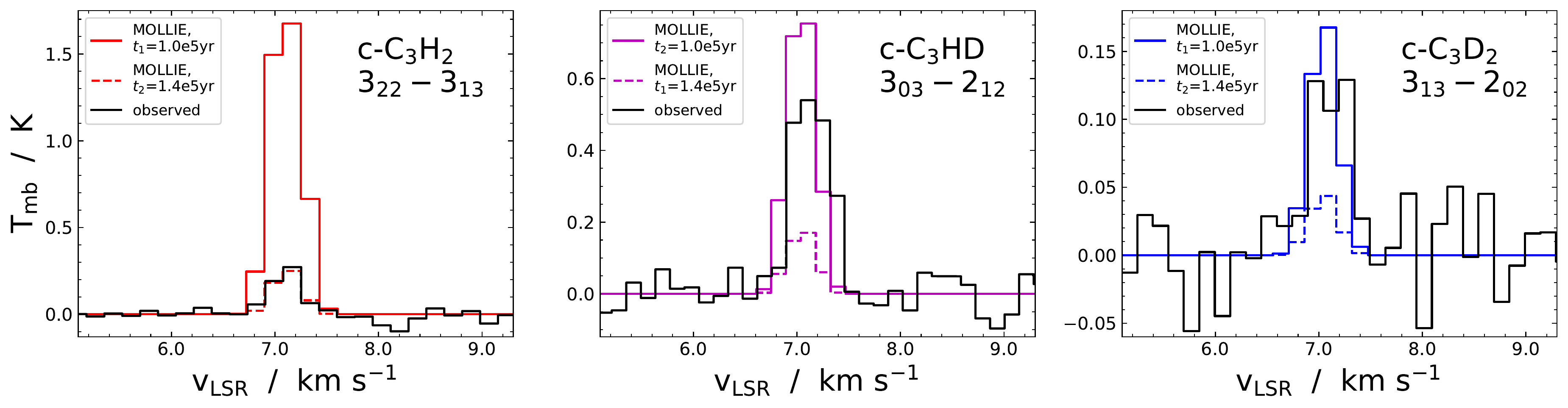}
   \caption{Comparison of the best fit solutions of spectra (coloured) modelled with MOLLIE at two different time steps of the chemical model, $t_1$\,=\,1.0$\times$10$^5$\,yrs (solid) and $t_2$\,=\,1.4$\times$10$^5$\,yrs (dashed), with the observed spectra (black). Plotted are the synthetic spectra of c-C$_3$H$_2$ (left), c-C$_3$HD (middle), and c-C$_3$D$_2$ (right). }
              \label{FigMollieResults}
\end{figure*}

\begin{table*}
    \centering
    \caption[Comparison of total column densities]{Comparison of total column densities, in units of $\times10^{12}$\,cm$^{-2}$}
    \begin{tabular}{lcccc}
    \hline
    \noalign{\smallskip}
     Molecule        & $N$(RADEX) & $N_{t_1}$(MOLLIE)  & $N_{t_2}$(MOLLIE) \\
     \noalign{\smallskip}
     \hline
     \noalign{\smallskip}
     c-C$_3$H$_2$   & 14(4)  & 160   & 30  \\
     c-C$_3$HD      & 3.8(9) & 3.8   & 0.8 \\
     c-C$_3$D$_2$   & 1.8(3) & 0.79  & 0.23\\
     \noalign{\smallskip}
    \hline 
    \noalign{\smallskip}
    \end{tabular}
    \tablefoot{The column densities derived with MOLLIE (N$_M$) are given at the chemical times $t_1$\,=\,1.0$\times$10$^5$\,yrs and $t_2$\,=\,1.4$\times$10$^5$\,yrs. The best fit for the deuterated species is $t_1$, for the main species $t_2$. The error on the derived values lies within a factor of 2. }
    \label{tab:compareNcol}
\end{table*}

\section{Discussion}\label{discussion}

The deuterium fraction maps presented in this work suggest that the deuteration of c-C$_3$H$_2$ is more efficient towards the centre of L1544. This implies that there is still c-C$_3$H$_2$ available to be deuterated also at high densities, despite the freeze-out. 
However, carbon chains are not expected to trace the dense central regions of pre-stellar cores, as they are expected to be frozen out. The detection of c-C$_3$D$_2$ peaking close to the dust peak demonstrates that a significant fraction of this species is still present towards the core centre. 
This might be due to the large amount of CO freeze-out,
as our chemical models show that with increasing CO freeze-out, the formation of He$^+$ is enhanced. With increasing level of He$^+$ in the gas phase, also the amount of carbon atoms needed to produce carbon chains increases, as He$^+$ efficiently destroys the CO molecules left in the gas phase via the reaction $\mathrm{He}^+ + \mathrm{CO} = \mathrm{C}^+ + \mathrm{O} + \mathrm{He}$.
This was already predicted by \cite{Ruffle1999}, whose models show an increase of C$^+$ at $3\times10^6$\,yrs, when CO starts its dramatic freeze-out, corresponding to a plateau in the HC$_3$N abundance (Figure 1 and 2 in \citealt{Ruffle1999})

The deuteration process of c-C$_3$H$_2$ consists of two reactions, the proton-deuteron transfer with H$_2$D$^+$ and the subsequent dissociative recombination with electrons (see \citealt{Spezzano2013}).
Therefore, the D/H ratio of c-C$_3$H$_2$ is directly related to that of H$_3^+$. The distribution of H$_2$D$^+$ across L1544 was shown to be extended and an excellent tracer of the dust continuum \citep{Vastel2006,Koumpia2020}, with the emission peak at the dust peak. The correlation between the degree of deuteration in molecules and the abundance of H$_2$D$^+$, found by \cite{Vastel2006}, is now further confirmed by our map of c-C$_3$D$_2$.

The deuterium fraction maps of c-C$_3$H$_2$ reveal how the efficiency of the deuteration varies across the core. However, it has to be taken into account that the deuterated isotopologues are much less abundant at larger radii than the main species, and more concentrated at higher densities (see abundance profiles in Fig.~\ref{Fig6Abundances}). Therefore, the column density of the normal species also contains contributions from the outer layers of the core, where c-C$_3$H$_2$ is still abundant, while c-C$_3$HD and c-C$_3$D$_2$ are not present. 
Due to this, the observed emission of the normal isotopologue does not necessarily come from the same layers where the deuterated isotopologues are located and the corresponding deuterium fraction maps have to be interpreted with caution. 
By analysing $R_{\rm D2/HD}$(c-C$_3$H$_2$), this problem is avoided, as the two molecules should trace more similar regions of the core (as also shown by the abundace profiles in Figure~\ref{Fig6Abundances}). 
The $R_{\rm D2/HD}$(c-C$_3$H$_2$) map shows a peak slightly shifted to the south-east of the dust peak, in between the dust and c-C$_3$H$_2$ peaks. This confirms that the deuteration process, and especially higher order deuteration, is more efficient towards the centre of the core.
The peak values of the deuteration maps derived in this work are $R_\mathrm{HD/H2}($c-C$_3$H$_2)=0.17$, $R_{\rm D2/H2}$(c-C$_3$H$_2)=0.025$ and $R_{\rm D2/HD}$(c-C$_3$H$_2)=0.16$, in good agreement with previous measurements towards the dust peak of L1544 (see e.g. \citealt{Spezzano2013}, \citealt{Chantzos2018}).

Using the non-LTE models RADEX and MOLLIE, we are able to reproduce the observed spectra towards the dust peak of L1544 in both cases. 
The derived column densities agree with the LTE calculations within a factor of 2. 
This shows that for our observed transitions, we can safely use RADEX for non-LTE calculations instead of a costly and time-consuming simulation like MOLLIE.
We further used the results of RADEX and MOLLIE to estimate the optical depth of the possibly optically thick transition of c-C$_3$H$_2$, ($3_{22}-3_{13}$). We found values ranging from 0.07 up to 0.3,
indicating optically thin emission.
Nevertheless, the non-LTE simulations confirm that the assumption of CTex is a useful approximation to derive the column density of optically thin lines. 

Recently, \cite{Redaelli2019} and \cite{ChaconTanarro2019} studied the deuteration throughout L1544, using emission maps of the molecules N$_2$H$^+$, HCO$^+$ and CH$_3$OH, H$_2$CO, respectively. 
The molecules discussed in \cite{Redaelli2019}, N$_2$H$^+$ and HCO$^+$, show a similar behaviour to c-C$_3$H$_2$: the D-bearing species peak closer to the dust peak than their corresponding normal species, and the deuteration maps show a compact morphology around the dust peak. 
N$_2$H$^+$ is more deuterated than c-C$_3$H$_2$, as N$_2$, and therefore N$_2$H$^+$, is a late-type molecule and suffers less of depletion in the dense central parts of the core. 
Thus, N$_2$H$^+$ reaches higher levels of deuteration (26\%) than c-C$_3$H$_2$ (17\%), tracing the densest regions. 
HCO$^+$, on the other hand, is formed from CO, which is highly depleted in the centre, so it traces outer layers of the core, similar to c-C$_3$H$_2$, only showing low levels of deuteration ($\approx 3.5$\%).

Using methanol (CH$_3$OH) and formaldehyde (H$_2$CO), which form on dust grain surfaces, \cite{ChaconTanarro2019} focus on the deuteration process in the solid phase. 
A comparison of the deuteration maps of methanol to the $R_{\rm HD/H2}$(c-C$_3$H$_2$) map and the $R_{\rm D2/H2}$(c-C$_3$H$_2$) map shows an interesting similarity. The deuteration peaks of both molecules are shifted from the dust peak towards the opposite direction of their respective molecular peaks, with respect to the major axis of L1544.
Following this, the morphology of the maps might not be the consequence of an increased deuteration, but rather a steep decrease of the column density of the main species in the outer layers of the core. The emission of the deuterated isotopologues comes instead only from the inner layers, and is not affected by the inhomogeneous interstellar radiation field around the core (that drives the segregation among methanol and c-C$_3$H$_2$). This is confirmed by the fact that the $R_{\rm D2/HD}$(c-C$_3$H$_2$) ratio peaks just next to the dust peak. 
The distribution of the deuterium fraction of H$_2$CO shows an opposite behaviour with respect to c-C$_3$H$_2$. The $R_{\rm HD/H2}$(c-C$_3$H$_2$) map is extended through the core, while for $R_{\rm HD/H2}$($\rm H_2CO$) the map is more centred around its peak located towards the north-west of the dust peak. For the double-deuteration maps the same applies, however, this time the distribution of $\rm H_2CO$ is more extended and c-C$_3$H$_2$ is more centered. 
The more extended $R_{\rm D2/H2}$($\rm H_2CO$) map might be linked to the fact that the $\rm H_2CO$ isotopologues have both grain-surface and gas-phase formation mechanisms, while c-C$_3$H$_2$ isotopologues solely form in the gas-phase.
The $R_{\rm D2/HD}$($\rm H_2CO$) map is not discussed in \cite{ChaconTanarro2019} due to high uncertainties and hence can not be used for a comparison.
A major difference between c-C$_3$H$_2$ and H$_2$CO is their deuteration level. For H$_2$CO, both $R_{\rm HD/H2}$($\rm H_2CO$) and $R_{\rm D2/H2}$($\rm H_2CO$) lie below 0.05.
Contrary to this, for c-C$_3$H$_2$, the singly deuterated species is much more abundant than the doubly deuterated one, reaching $R_{\rm HD/H2}$(c-C$_3$H$_2)=0.17$ and $R_{\rm D2/H2}$(c-C$_3$H$_2)=0.025$, respectively. This indicates that for $\rm H_2CO$, higher order deuteration happens much more efficiently.

\begin{figure}
   \includegraphics[width=\hsize]{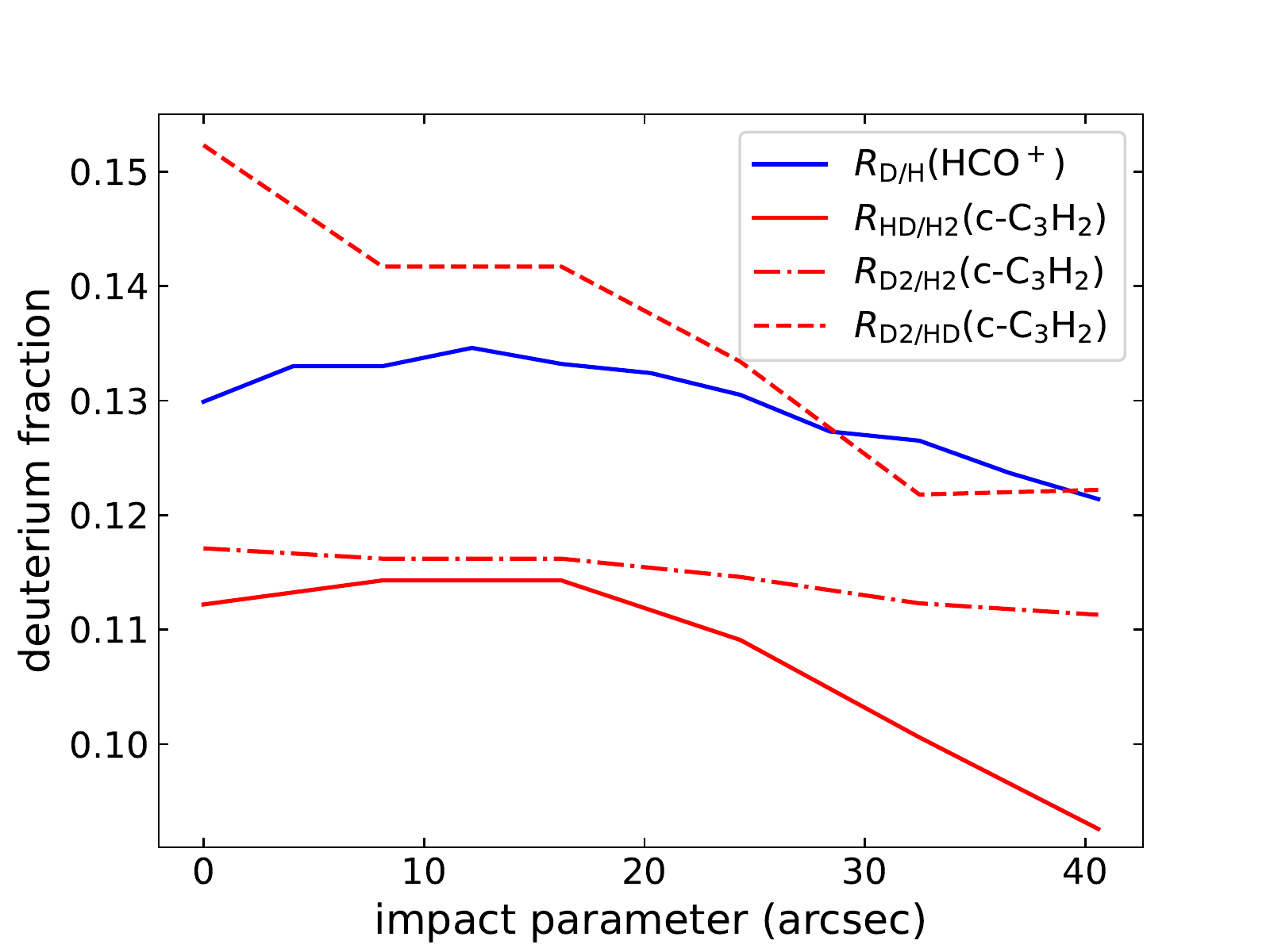}
   \caption{Comparison of the trends of
    the HCO$^+$ deuteration map presented in \cite{Redaelli2019} and the deuteration maps of c-C$_3$H$_2$ derived in this work,
    along the cut shown in Figure 10 in \cite{Redaelli2019}. The data points of $R_{\rm HD/H2}(\mathrm{HCO}^+)$ and $R_{\rm D2/H2}$(c-C$_3$H$_2)$ are shifted upwards by 0.10 to allow an easier comparison.}
              \label{Figdhcut}
\end{figure}

By cutting across the dust peak and the $R_{\rm HD/H2}(\mathrm{HCO}^+)$ maximum,
\cite{Redaelli2019} compared the trends of the deuteration maps of their molecular tracers. Figure \ref{Figdhcut} applies the same cut to the deuteration maps of c-C$_3$H$_2$ derived in this work, plotting the deuteration fractions as a function of the impact parameter (=projected distance to the dust peak) and comparing them to HCO$^+$. The cut is visualized as a dashed line in the left panel of Fig.~\ref{FigDeutFrac}. Along the cut, the deuteration levels of $R_{\rm HD/H2}$(c-C$_3$H$_2)$ and $R_{\rm D2/HD}$(c-C$_3$H$_2$) decrease faster than $R_{\rm D2/H2}$(c-C$_3$H$_2)$ and $R_{\rm HD/H2}(\mathrm{HCO}^+)$. 
The slope of $R_{\rm HD/H2}$(c-C$_3$H$_2)$ shows a similar behaviour as HCO$^+$, as both peak between 10$"$-20$"$ from the dust peak and then drop off, though the one of c-C$_3$H$_2$ is steeper.
The fast decrease of $R_{\rm D2/HD}$(c-C$_3$H$_2$) with increasing distance from the dust peak confirms once more that the deuteration process is most efficient in the central region. This enhanced c-C$_3$H$_2$ deuteration towards the centre of a pre-stellar core could explain the large $R_{\rm HD/H2}$(c-C$_3$H$_2)$ values found by \cite{Chantzos2018} towards the protostar HH211, one of the youngest class 0 sources.

\FloatBarrier
\section{Conclusions}\label{conclusion}

In this work, we studied the deuteration process of cyclopropenylidene towards the pre-stellar core L1544. 
We presented the first emission map of doubly deuterated c-C$_3$H$_2$, along with the emission maps of its singly deuterated and non-deuterated isotopologues. Peaking at the centre of the core, the column density map of c-C$_3$D$_2$ proves that carbon-chain molecules are still present at the high central densities. This is possible due to an increased abundance of He$^+$ that destroys CO molecules and therefore increases the amount of free carbon atoms available to form c-C$_3$H$_2$ \citep{Ruffle1999}.
Furthermore, the distributions of c-C$_3$HD and c-C$_3$D$_2$ indicate that the deuterated forms of c-C$_3$H$_2$ are preferentially located towards the centre of L1544, where the increased abundance of H$_2$D$^+$ drives the deuteration process, and do not trace the c-C$_3$H$_2$ peak located in the south-east, which is caused by external illumination of the interstellar radiation field.
Using the column density maps calculated assuming CTex, we derived the first deuteration maps of c-C$_3$H$_2$. Showing a maximum at the dust peak, the maps confirm that the deuteration process of c-C$_3$H$_2$ is more efficient towards the centre of the core. The maximum values in the maps are $R_{\rm HD/H2}$(c-C$_3$H$_2)=0.17\pm0.01$, $R_{\rm D2/H2}$(c-C$_3$H$_2)=0.025\pm0.003$ and $R_{\rm D2/HD}$(c-C$_3$H$_2)=0.16\pm0.03$, which are in good agreement with previous single point measurements \citep{Spezzano2013,Chantzos2018}. 

Using the non-LTE radiative transfer models RADEX and MOLLIE, we were able to reproduce the spectra of c-C$_3$H$_2$, c-C$_3$HD and c-C$_3$D$_2$ at the dust peak of L1544.  The derived non-LTE column densities agree with the CTex calculations within a factor of 2. The comparison of our methods shows that for our observed transitions, we can safely use RADEX for non-LTE calculations instead of a time-consuming simulation like MOLLIE.

Finally, we compared the deuterium fraction maps derived in this work to the deuteration of other molecular tracers. The discussion showed how studying the deuteration of different species can bring important information on the chemistry of different parts of the source and helps to build a comprehensive picture of L1544.
However, so far only a limited number of deuteration maps is available in the literature. Future projects will be able to address this deficit and lead to a deeper understanding of the deuteration process of individual molecules.
In particular, future studies can further investigate the detection of carbon-chain molecules at unexpectedly high densities to put tighter constraints on gas-grain chemical models. Suitable candidates for observations of other abundant carbon-chains are for example CH$_3$CCH and HC$_3$N, which show a not yet understood segregation towards the pre-stellar core L1544 (see \citealt{Spezzano2017}).

\begin{acknowledgements}
    S.S.  and K.G. wish to thank the Max Planck Society for the Max Planck Research Group funding. All others authors affiliated to the MPE wish to thank the Max Planck Society for financial support.
\end{acknowledgements}

%
\bibliographystyle{aa} 
\bibliography{mybib.bib} 
%

\begin{appendix}

\FloatBarrier
\section{Column density maps}\label{section:ncolmaps}
\begin{figure*}[h]
   \centering
   \includegraphics[width=.48\textwidth]{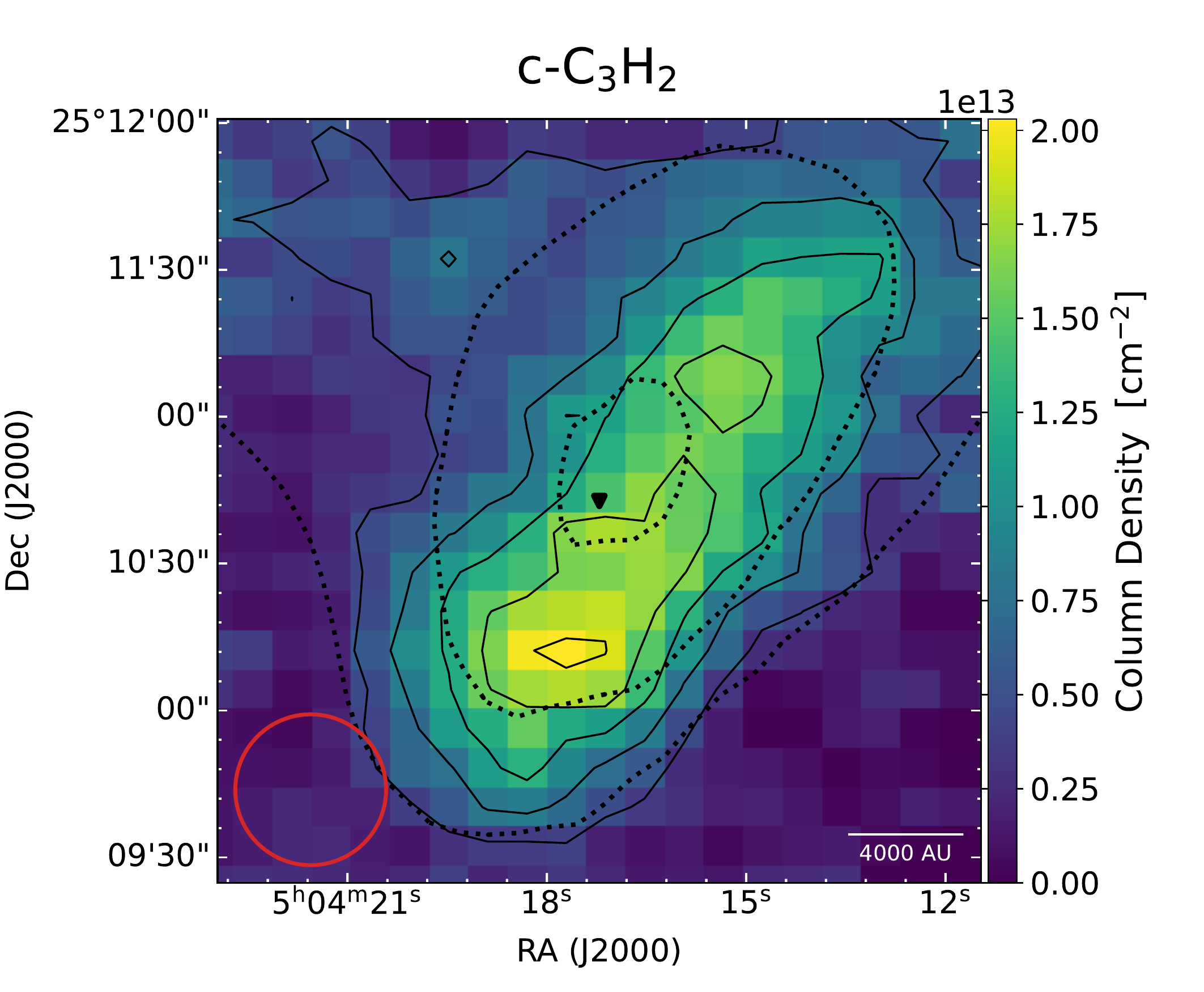}
   \includegraphics[width=.48\textwidth]{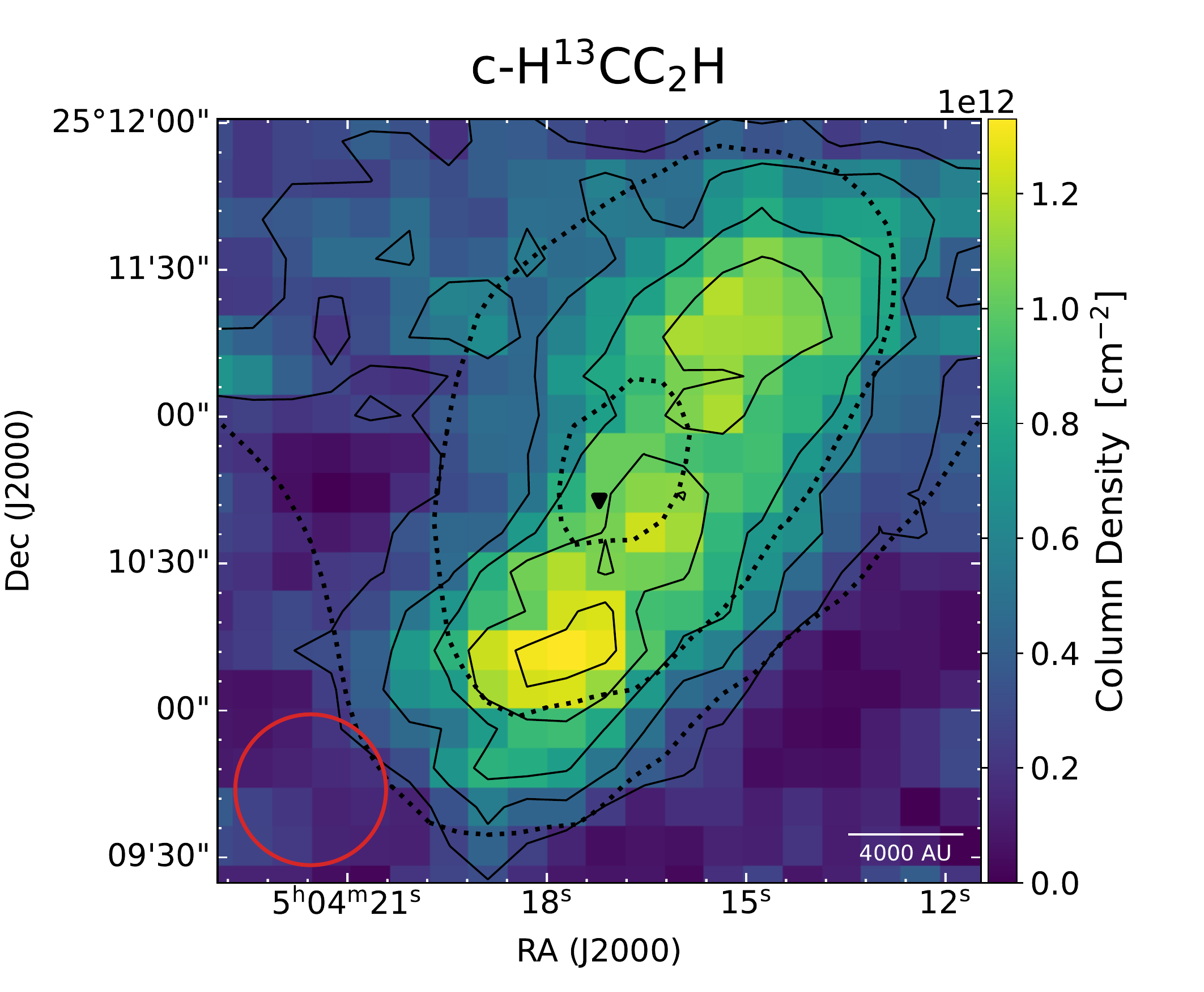}
   \includegraphics[width=.48\textwidth]{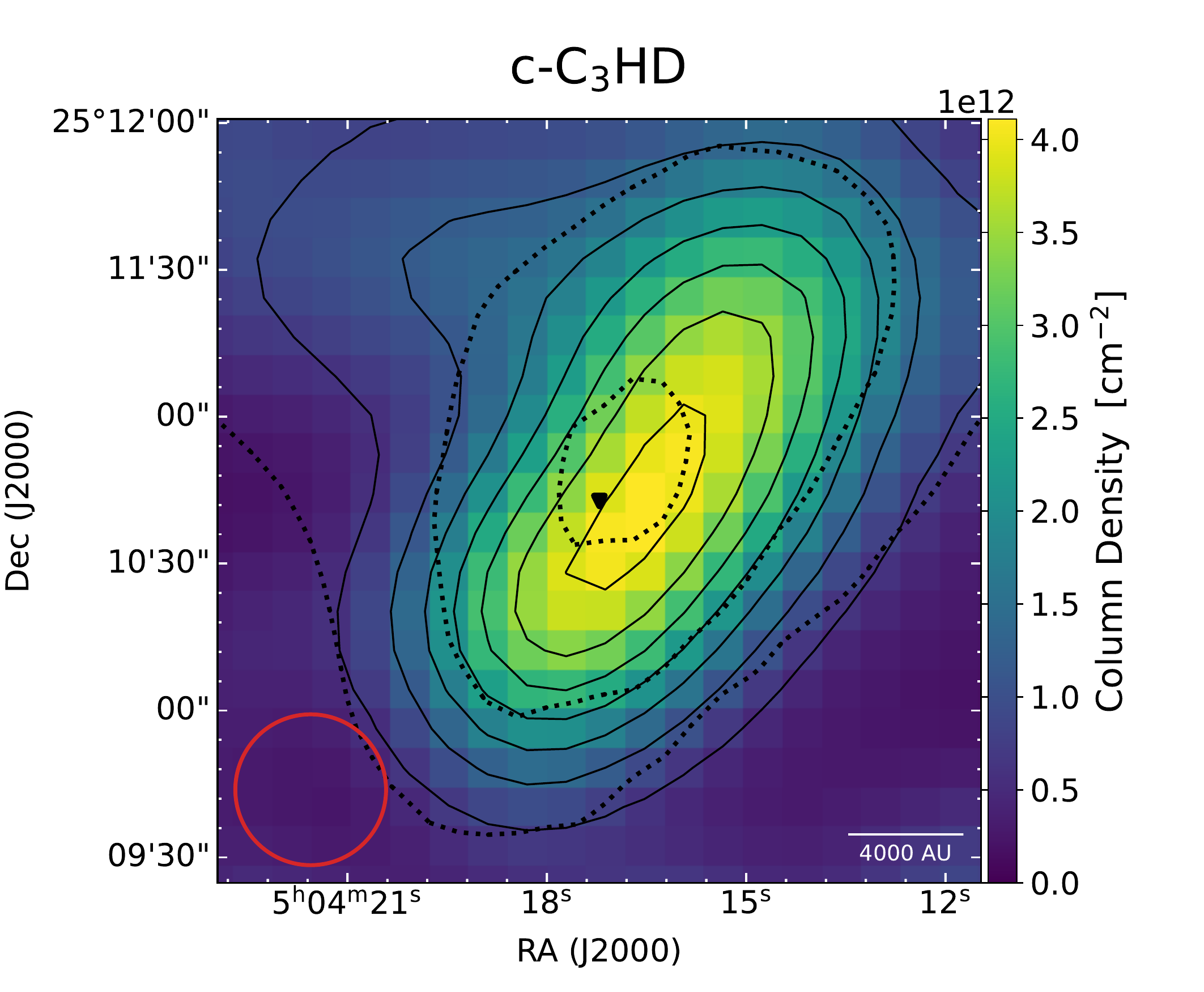}
   \includegraphics[width=.48\textwidth]{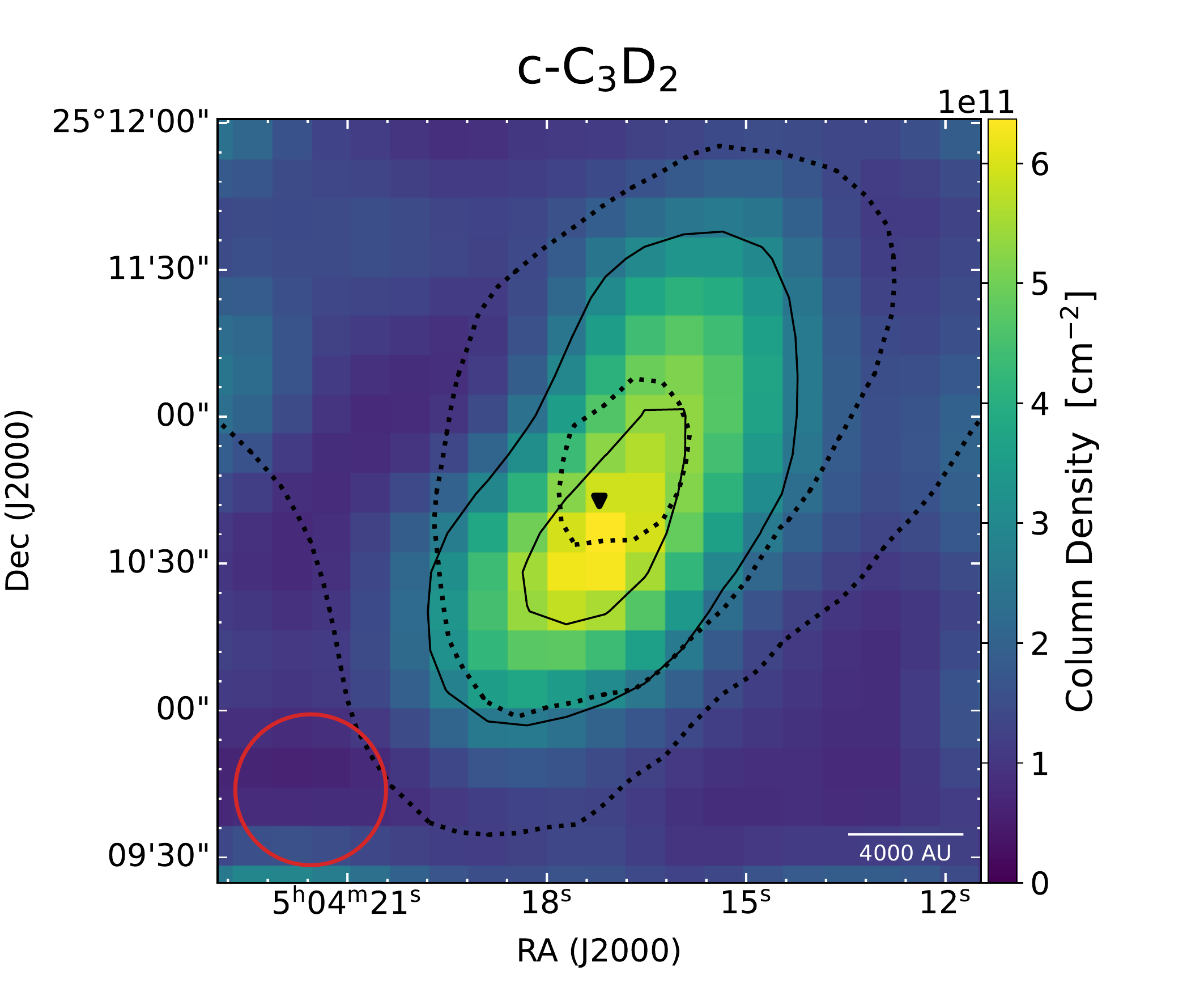}
   \caption{Column density maps of the molecules, used to calculate the deuteration fraction maps. The black dashed lines represent the 90\%, 50\% and 30\% of the H$_2$ column density peak value derived from \textit{Herschel} maps \citep{Spezzano2016b}, 2.8$\times$10$^{22}$\,cm$^{-2}$. The solid lines represent contours of the column density, starting at $3\sigma$ with steps of $3\sigma$. From top to bottom and left to right, the average errors on the column density are 1$\times10^{12}$, 9$\times10^{10}$, 2$\times10^{11}$ and 1$\times10^{11}$, in units of cm$^{-2}$. The dust peak is indicated by the black triangle. The beam size of the 30\,m telescope, $\rm HPBW=31''$, is shown by the red circle, and the scalebar is shown in the bottom right corners.}
              \label{FigNcolmaps}
\end{figure*}

In Figure \ref{FigNcolmaps} we show the derived column density maps of the observed molecules, using the parameters listed in Table \ref{TabNcolParam}.

\vfill

\FloatBarrier
\section{Averaged spectra}\label{section:averagedspectra}
\begin{figure*}
        \centering
        \includegraphics[width=0.99\textwidth]{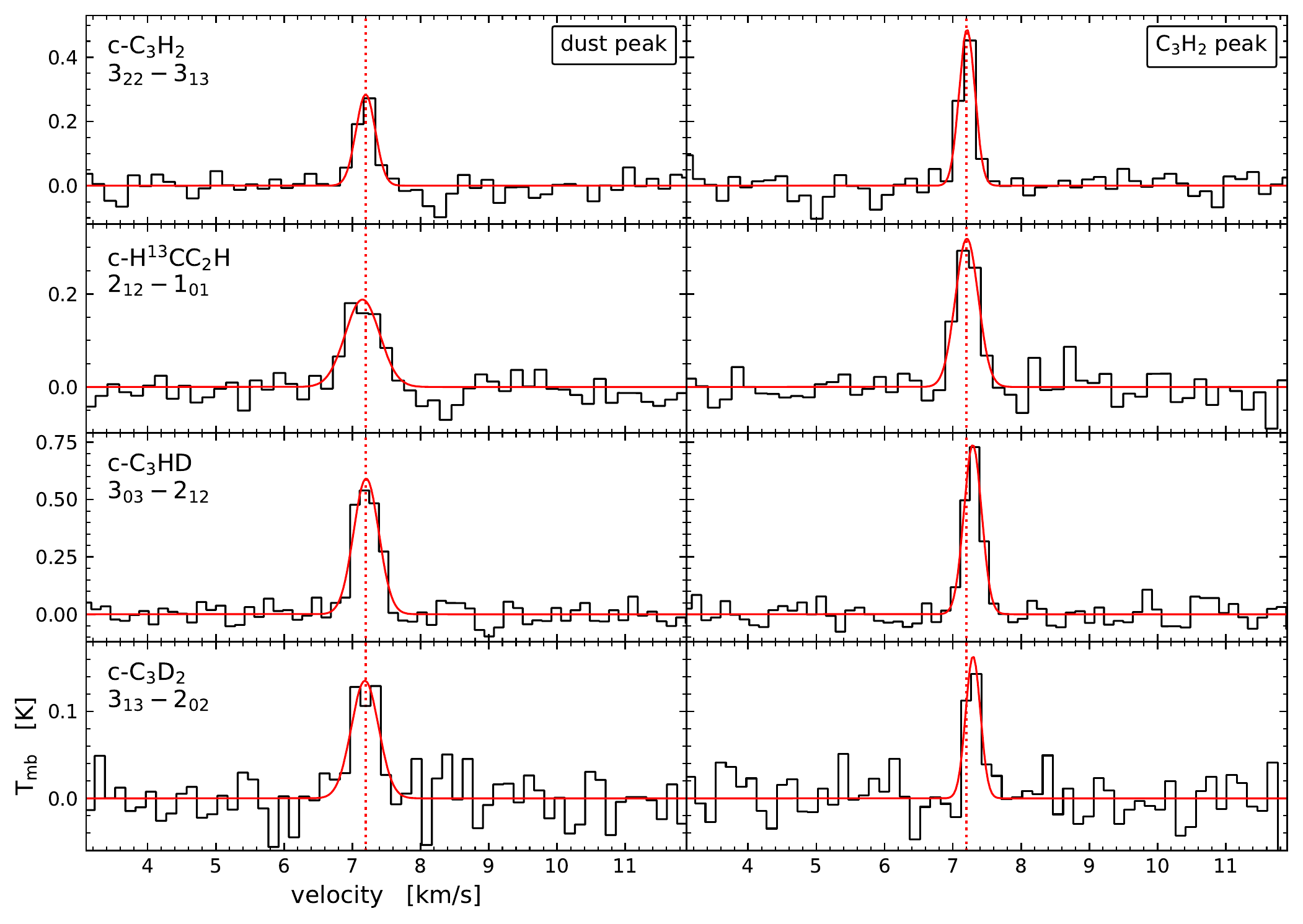}
        \caption{Comparison of extracted spectra (black) of the transitions listed in Table \ref{LineParam}, at the dust peak (left) and the c-C$_3$H$_2$ peak (right) of L1544. Overlaid are the fitted Gaussian profiles (red). The vertical lines indicate the systemic velocity of the source, $7.2$\,km\,s$^{-1}$.}
        \label{FigAveSpectra}
\end{figure*}

Figure \ref{FigAveSpectra} shows the spectra extracted at the dust peak \citep{WardThompson1999} and the c-C$_3$H$_2$ peak \citep{Spezzano2016b} of L1544. Overlaid are fitted Gaussian profiles (red).

\FloatBarrier
\section{Variation of input parameters of the chemical model}\label{section:testchemicalmodel}
\begin{figure*}
        \centering
        \includegraphics[width=0.8\textwidth]{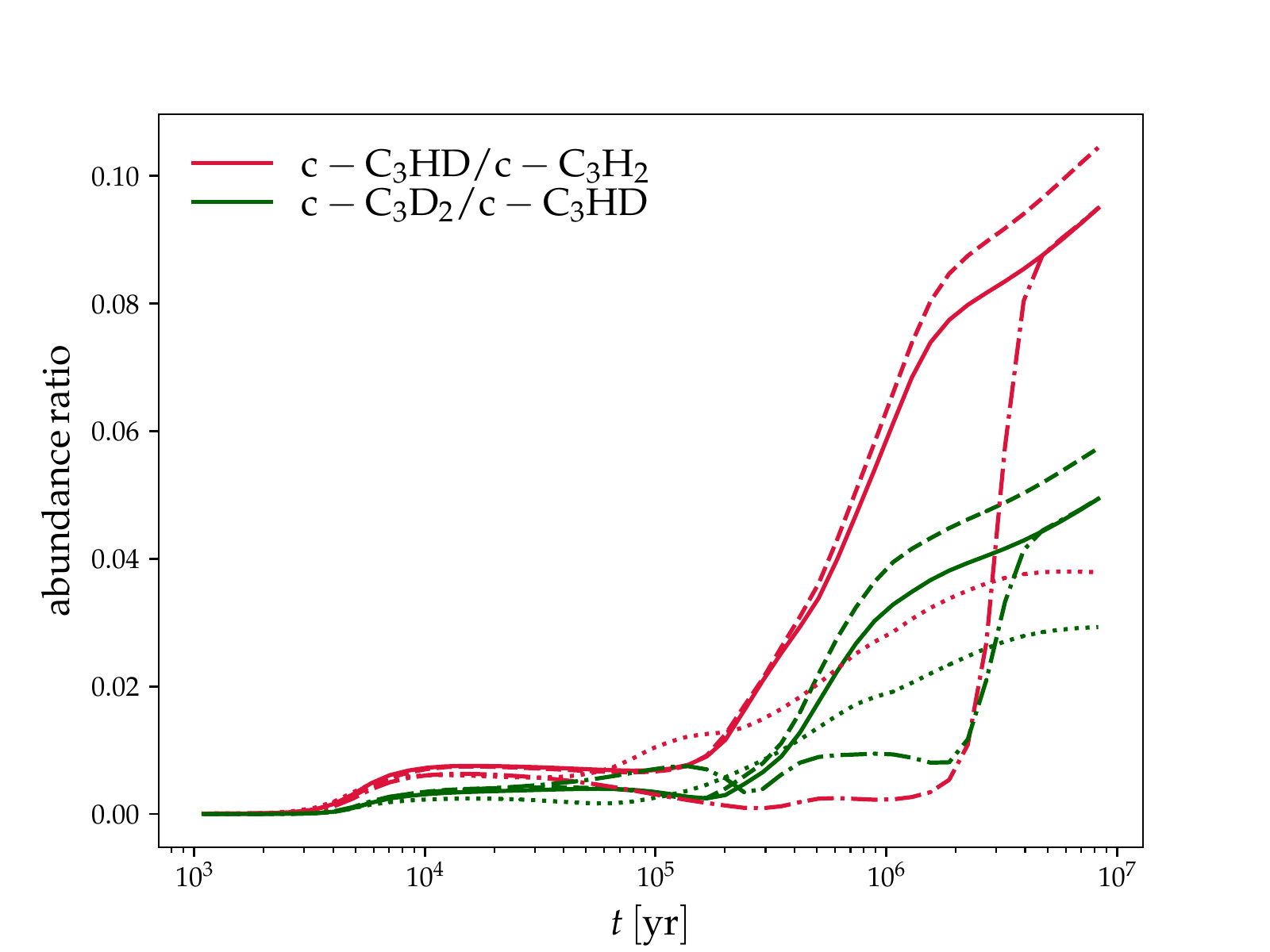}
        \caption{Comparison of the evolution of the D/H ratios of c-C$_3$H$_2$ with varied input parameters in the chemical model: fiducial model (solid lines), decreased C/O ratio (dashed lines), higher CR ionization rate (dotted lines), higher initial H$_2$ ortho-to-para ratio (dashed-dotted lines).}
        \label{FigChemicalModelTest}
\end{figure*}

To test the effect of the various parameters, we run four single-point simulations, assuming physical conditions roughly representative of the outer core region where the c-C$_3$H$_2$ abundance peaks: $n(H_2) = 10^4$\,cm$^{-3}$, $T=10$\,K, $A_\mathrm{V} = 5$\,mag. In addition to the fiducial parameter set, we test three variations with respect to the fiducial model: 1) decreased C/O ratio (from $\sim$.47 to $\sim$.35); 2) higher cosmic-ray ionization rate ($\zeta=10^{-16}$\,s$^{-1}$); 3) higher initial H$_2$ ortho-to-para ratio (3).
Figure~\ref{FigChemicalModelTest} shows the resulting D/H ratios as a function of time in these four cases. The solid lines represent the fiducial case, while the dashed, dotted, and dash-dotted lines represent cases 1 to 3, respectively, as defined above.

A variation of the applied C/O ratio from $\sim$0.47 down to $\sim$0.35 only slightly changes the results. An increase of the initial ortho-to-para ratio of H$_2$ up to o:p\,$=$\,3 simply delays the evolution of the D/H ratios, eventually arriving at the same end result as the in the case of a low initial ratio. This behaviour was also observed by \cite{Kong2015}. Only the cosmic-ray ionization rate seems to have a larger effect, where an increase up to 10$^{-16}$\,s$^{-1}$ reduces both D/H ratios. Previous work done by \cite{Redaelli2021} found the ionization rate to be $\sim3\times10^{-17}$\,s$^{-1}$, which is less than a factor of 3 different from our fiducial value.
The test simulations show that the time is not the only variable that influences the D/H ratios of c-C$_3$H$_2$. While we only explored a small part of the paramter space, and the tests were limited to a single point and cannot be used to conclude anything on the column densities, we do not expect variations of more than a factor of a few in the c-C$_3$H$_2$ D/H ratios, as long as the model parameters remain reasonably close to those in the fiducial model. A complete parameter-space explorations is out of the scope of the present work.

\end{appendix}

\end{document}